\documentclass[usenatbib]{mn2e}

\pdfoutput=1

\usepackage[pdftex]{graphicx} 

\usepackage{times}

\usepackage{rotate} 

\usepackage{amssymb} 

\usepackage{rotating}

\usepackage{epstopdf} 

\usepackage{graphicx, subfigure}

\usepackage{array}

\usepackage{booktabs}

\title[Searching for luminous absorbed sources in the WISE AGN catalogue]{Searching for luminous absorbed sources in the WISE AGN catalogue}

\author[Mountrichas et al.]{G. Mountrichas$^1$,
  I. Georgantopoulos$^{1}$, N. J. Secrest$^{3}$, I. Ordov\'as-Pascual$^{2}$,  A. Corral$^{1}$,
  \newauthor  A. Akylas$^{1}$, Mateos, S$^{2}$, Carrera, F. J.$^{2}$, Batziou, E. $^{4}$\\ \\ 
$^1$National Observatory of Athens, V.  Paulou  \& I.  Metaxa, 11532,  Greece\\
$^2$Instituto de F�sica de Cantabria (CSIC-UC), Avenida de los Castros, 39005 Santander, Spain\\
$^3$Department of Physics \& Astronomy, George Mason University, MS 3F3, 4400 University Drive, Fairfax, VA 22030, USA\\
$^4$Faculty of Physics, School of Sciences, Univ. of Athens, Panepistimiopolis, 15771 Ilissia, Greece}

\begin{document}
\maketitle
\label{firstpage}

\begin{abstract}  

Mid-IR colour selection techniques have proved to be very efficient in finding AGN. This is because the AGN heats the surrounding dust producing warm mid-IR  colours. Using the WISE 3.6, 4.5 and 12  $\mu m$ colours, the largest sample of IR selected AGN has already been produced containing 1.4 million AGN over the whole sky. Here, we explore the X-ray properties of this AGN sample by cross-correlating it with the subsample of the 3XMM X-ray catalogue that has available X-ray spectra and at the same time optical spectroscopy from SDSS. 
Our goal is to find rare luminous obscured AGN. Our final sample contains 65 QSOs with $\rm{log}\,\nu L_\nu \ge 46.2$\,erg\,s$^{-1}$. This IR luminosity cut corresponds to $\rm{log}\,L_X \approx 45$\,erg\,s$^{-1}$, at the median redshift of our sample ($z=2.3$), that lies at the bright end of the X-ray luminosity function at $z>2$. The X-ray spectroscopic analysis reveals seven obscured AGN
 having a column density $\rm N_H>10^{22} cm^{-2}$. Six of them show evidence for broad [CIV] absorption lines and five are classified as BALQSOs. We fit the optical spectra of our X-ray absorbed sources to estimate the optical reddening. We find that none of these show any obscuration according to the optical continuum. These sources add to the growing evidence for populations of luminous QSOs with evidence for substantial absorption by outflowing ionised material, similar to those expected to be emerging from their absorbing cocoons in the framework of AGN/galaxy co-evolution.
 

\end{abstract}


\section{Introduction}

Mid-IR selection techniques provide a very efficient means to detect Active Galactic Nuclei (AGN). 
These methods are based on the detection of the 'warm' colours arising from the hot dust surrounding the AGN. This dust, most probably in the form of a clumpy torus \citep{Wada2011, Nenkova2008a}, is heated by the central black hole radiation and is re-emitted in the mid-IR.  The mid-IR AGN surveys have been introduced by \cite{Lacy2004, Stern2005, Alonso2006, Donley2007} using data from the {\it Spitzer} mission \citep{Werner2004}. 
 These colour-selection techniques  have been adapted for use in the Wide-field Infrared Survey Explorer ({\it WISE}) mission \citep{Wright2010}. 
  {\it WISE} performed an all-sky survey in four mid-IR bands 3.4, 4.6, 12 and 22\,$\rm \mu m$. 
In particular, \cite{Stern2012} applied a one colour selection criterion using the 3.4 and 4.6 $\mu m$ bands \citep[see also][]{Assef2013}.  
\cite{Mateos2012} proposed a  'wedge' selection method based on three {\it WISE} colours. 
 All the above methods are particularly efficient in finding the most luminous QSOs. However, they become incomplete at lower 
luminosities \citep[see e.g][]{Barmby2006, Georgantopoulos2008, cardamone2008, Donley2012, Mateos2012}.
This is because contamination by the host galaxy renders the mid-IR colours bluer. 
 Recently, \cite{Secrest2015} applied the \cite{Mateos2012} selection criteria in the entire {\it WISE} dataset compiling a sample
of 1.4 million AGN candidates. 

The \cite{Secrest2015} catalogue is useful in finding the most luminous AGN and among them, those that are obscured. Luminous obscured AGN are quite rare. X-ray surveys have demonstrated that the fraction of obscured AGN, defined as those with a column density higher than $\rm N_H=10^{22}$\,cm$^{-2}$, drops significantly with increasing luminosity \citep{ueda2003, akylas2006, Della_ceca2008, Assef2013,Ueda2014}. This could naturally occur because the radiation pressure from a powerful AGN blows the 
torus away  \citep{Fabian2008}. According to models of galaxy formation \citep{Hopkins2008a}, the most luminous AGN 
may represent the product of galaxy mergers. Then obscured and luminous AGN represent the phase where the AGN is still growing 
and has not yet blown away its obscuring screen. \citet{Brusa2015} find a number of luminous obscured AGN in the COSMOS field which show strong  evidence for 
  interaction with the host galaxy in their optical spectra. 
 More interestingly, a few wide area surveys found a number of obscured sources  among optical type-1, broad-line, AGN. 
\cite{Page2001} selected absorbed sources in {\it ROSAT} pointed observations. These presented moderate amounts of obscuration ($N_H\sim 10^{22} \,$cm$^{-2}$) based on their hardness ratios (HRs). They were associated with 
 type-1 QSOs according to their optical spectral classification. \cite{Wilkes2005} investigated the X-ray properties of 2MASS selected red QSOs using both {\it Chandra} and {\it XMM-Newton}. These presented moderate amounts of obscuration in their X-ray spectra. Finally, previous studies \citep[e.g.,][]{Wolter2005, Mateos2005, Merloni2014} find that an appreciable fraction (20-30\%) of optical type-1 AGN present obscuration in X-rays. These sources may contain more gas than dust and could represent the phase where the AGN is blowing away the obscuring cocoon. 

In this paper, we expand the above works by studying the X-ray properties of  {\it WISE} selected AGN. Our main goal is to find rare luminous obscured AGN. Towards this end, we cross-correlate the {\it WISE} AGN sample of \cite{Secrest2015} with the 3XMM catalogue \citep{Rosen2016}. The {\it WISE} AGN survey, owing to its large cosmological volume provides  a unique database for the study of this rare class of AGN.  We focus our analysis on the 3XMM-DR5 catalogue sub-sample with available X-ray spectroscopy \citep{Corral2015} and optical spectra from the  Data Release 12 (DR12) of SDSS/BOSS \citep{Alam2015}. The sources found to be absorbed, based on their X-ray spectra, are also investigated for possible absorption features in their optical spectra. Moreover, reddening estimates from 
the optical spectra as well as from the optical to mid-IR colours are presented. 

We adopt H$_0$ = 75\,km\,s$^{-1}$\,Mpc$^{-1}$, $\Omega_M$ = 0.3 and $\Omega_\Lambda$ = 0.7 throughout the paper.

\section{Data}

The infrared (IR) and X-ray catalogues used in this analysis, as well as our final AGN sample constructed by the cross-matching of the two, are described in the following.

\subsection{The infrared sample}
The infrared AGN catalogue used in our analysis, was extracted from the WISE data, as described in \cite{Secrest2015}. The WISE survey is an all-sky mid-IR survey at four bands, i.e., 3.4, 4.6, 12 and 22 $\mu$m \citep[W1, W2, W3, W4, respectively][]{Wright2010}. The AllWISE data set, that was used for the AGN extraction, contains positions, apparent motions, magnitudes and PSF-profile fit information for almost 748 million objects. 

A large number of AGN mid-IR colour selection criteria exists in the literature. These colour criteria rely on the fact that in mid-IR colour space, AGN separate well from stars and star-forming galaxies. Nevertheless, a number of requirements have to be met, in choosing those colour criteria that can be applied on the WISE sample. The criteria have to be defined directly from the WISE dataset to avoid  uncertainties related to the transformation from different magnitude systems. \cite{Secrest2015} used the W1, W2 and W3 bands of WISE, along with equations (3) and (4) from \cite{Mateos2012}  to define the WISE AGN sample. The selection of W2 and W3 bands also reduces the bias against heavily absorbed AGN and minimizes the contamination by non-AGN sources, i.e., star-forming galaxies \citep{Mateos2012}. The final sample consists of 1,354,775 mid-IR AGN candidates covering relatively uniformly the full sky. Approximately 1.1 million of these were previously uncatalogued.

\subsection{The X-ray sample}

The X-ray sample used is the 3XMM-Newton Data Release 5 \citep{Rosen2016} that contains 565,962 detections, covering 877\,deg$^2$.  396,910 of these sources are unique. The detections are drawn from 7781 XMM-Newton EPIC observations, within an energy range from 0.2\,keV to 12\,keV. 91,679 sources have available spectra. Spectra were considered for extraction for point sources with a count limit of 100 counts in EPIC. A detailed description of the 3XMM catalogue and the spectral extraction is available online \footnote{\textrm{http://xmm-catalog.irap.omp.eu/docs/spectral-fitting}}.

\subsection{Sample selection}

\begin{figure}
\begin{center}
\includegraphics[height=1.\columnwidth]{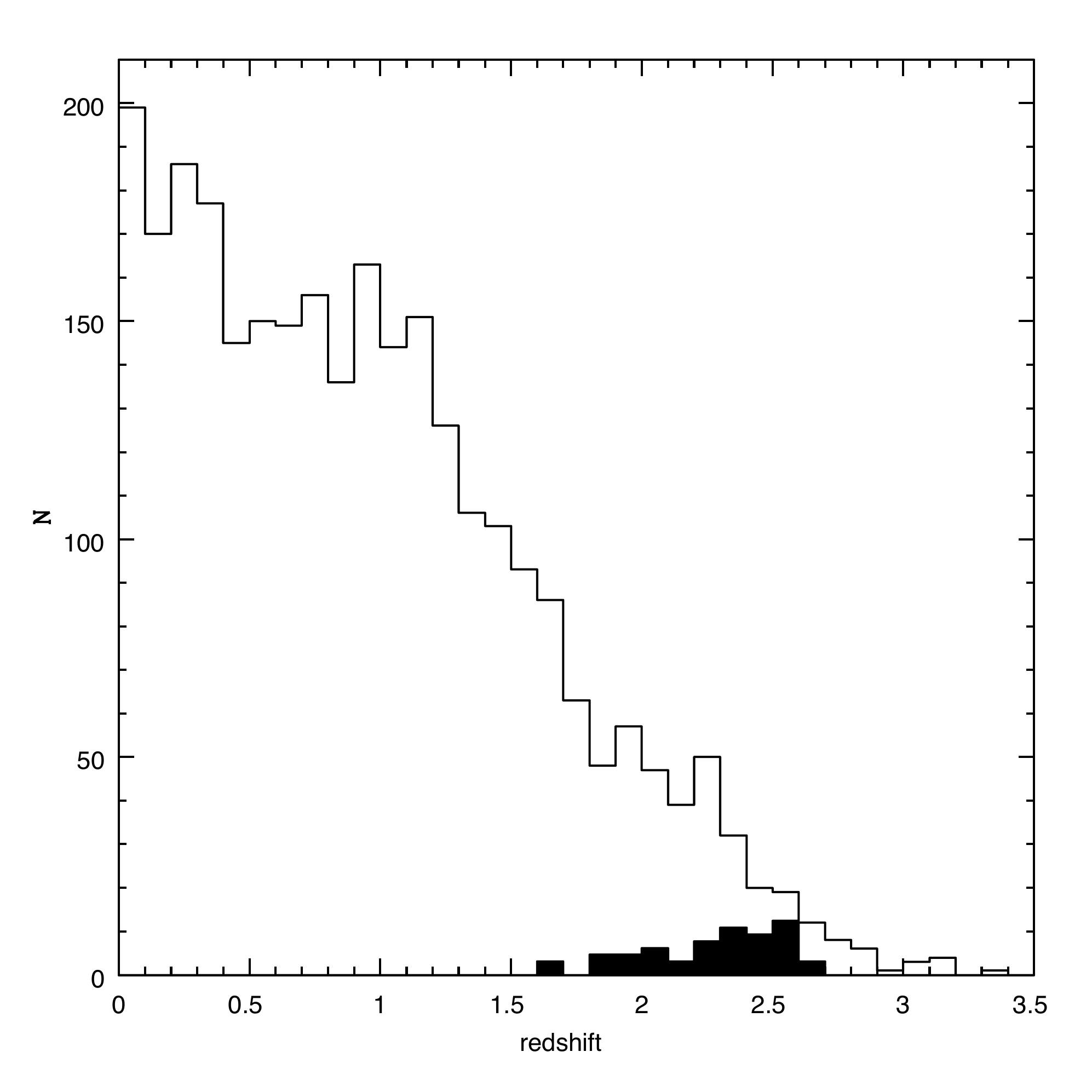}
\end{center}
\caption{The redshift distribution of the 2,834 AGN, from the cross-correlation of the WISE and 3XMM catalogues. The black shaded region, shows the redshift distribution of our luminous sample of 65 AGN (see text).}
\label{fig_n_z}
\end{figure}

\begin{figure*}
\begin{center}
\includegraphics[height=1.\columnwidth]{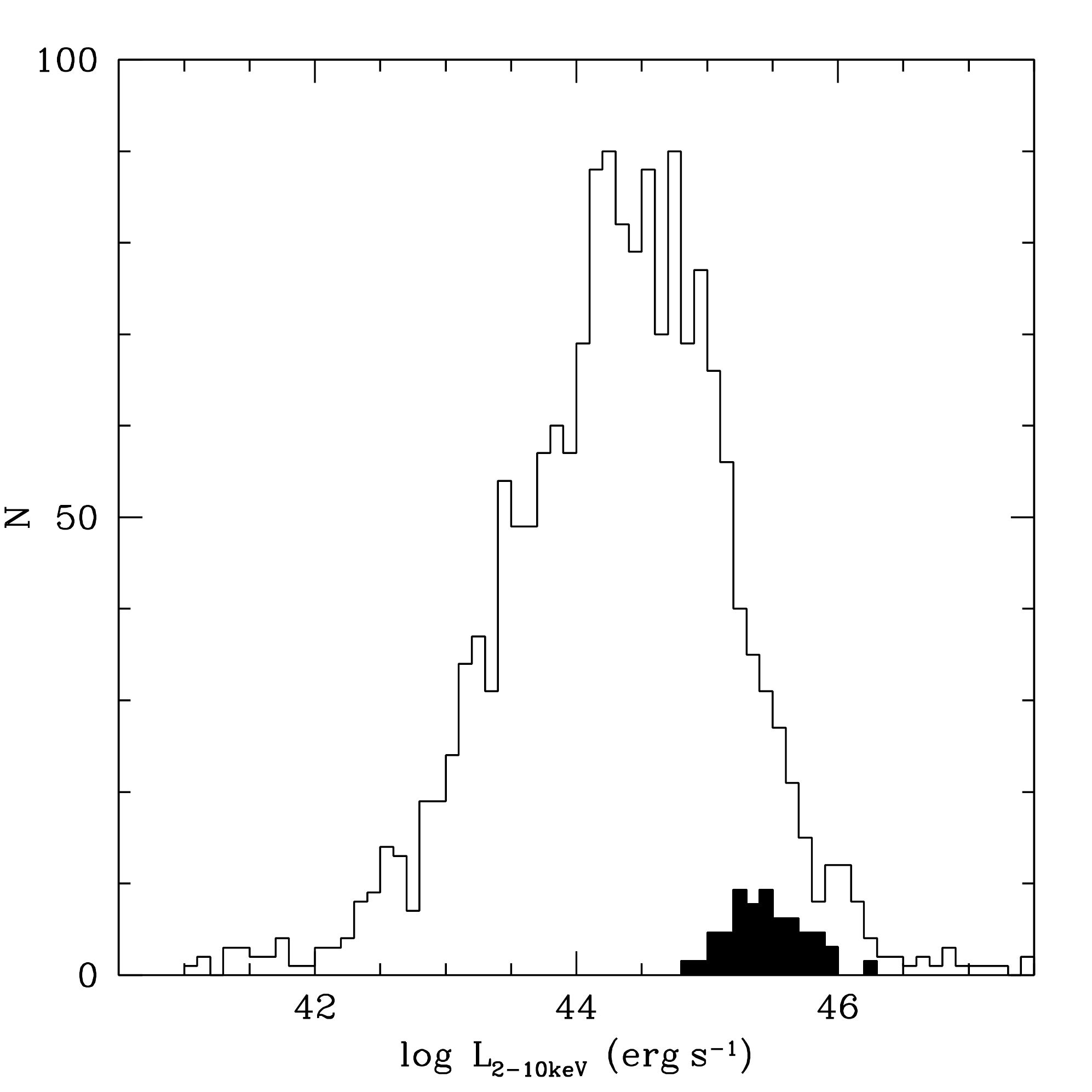}
\includegraphics[height=1.\columnwidth]{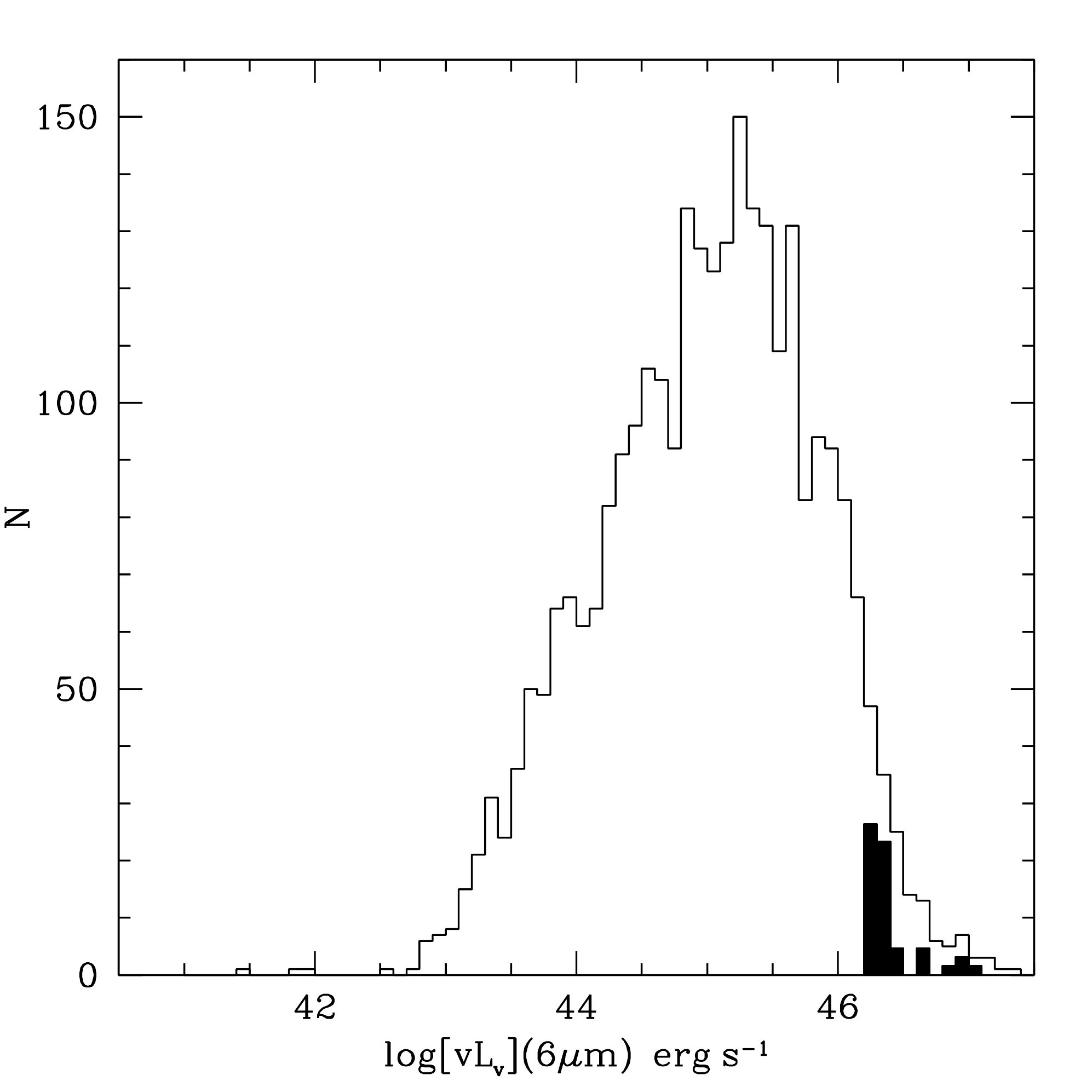}
\end{center}
\caption{Left:  The X-ray (observed), log$[L_{2-10keV}]\,$erg\,$s^{-1}$, luminosity distribution of the 2,834 AGN. Right: The infrared, log$[vL_v](6\mu m)$\,erg\,$s^{-1}$, luminosity distribution of the 2,834 AGN. The black shaded regions, show the luminosity distribution of our luminous sample of 65 AGN (see text).}
\label{fig_lum_distrib}
\end{figure*}

The scope of this work is to find AGN that are luminous based on their rest-frame, infrared luminosity, $\nu L_\nu$, and absorbed, based on the values of their column densities, $N_H$. We chose to use the infrared luminosity because it provides the most unbiased measurement of the bolometric luminosity compared to other wavelengths. 6$ \mu m$ is chosen as it is a good proxy of the torus luminosity and this spectral region is free of strong galaxy features \citep[e.g.,][]{Mateos2015}. An AGN sample that contains the infrared as well as the X-ray properties of the sources is needed. For that purpose, the two catalogues, described in the previous sections, are cross-correlated. For the cross-match we use TOPCAT, version 4.2-3\footnote{ http://www.star.bris.ac.uk/~mbt/topcat/}, using a radius of 3 arcsec. The position accuracy for a typical source in the DR5 of the 3XMM catalogue is 1 - 2 arcsec \citep{Rosen2016}, while in the WISE sample the precision is better than 1$''$ \citep{Wright2010}. 5,990 AGN are found to be common in the two catalogues and 2,834 of them also have available redshift. Redshifts come preferentially from LQAC-2 \citep{Souchay2012}, then from DR12 of SDSS and then MILLIQUAS \citep[for more information see][]{Secrest2015}. To check the fraction of expected spurious matches, each 3XMM source is re-positioned within a radius up to 5 arcmin around its original position. 100 random catalogues are created, following this process and then they are cross-matched with the WISE sample. The test reveals that we expect to have about $0.2\%$ of spurious matches in our sample. The redshift as well as the X-ray (observed) luminosity distributions, in the 2$-$10\,KeV band, of the 2,834 sources are presented in Figures  \ref{fig_n_z} and \ref{fig_lum_distrib}, respectively. 



To select luminous AGN, the $\nu$L$\nu$ at 6 $\mu$m is calculated. Contamination from star formation at  6 $\mu$m is negligible \citep[e.g][]{Symeonidis2016}, especially at the bright luminosities that our final AGN sample spans (see below). First, the VEGA magnitudes of the WISE catalogue are converted to AB, using the following expressions  \footnote{\textrm{http://wise2.ipac.caltech.edu/docs/release/prelim/expsup/}},

\begin{equation}
W1_{AB}=W1+2.683,
\end{equation}
\begin{equation}
W2_{AB}=W2+3.319,
\end{equation}
\begin{equation}
W3_{AB}=W3+5.242,
\end{equation}
\begin{equation}
W4_{AB}=W4+6.604.
\end{equation}

\noindent The observed wavelength, $\lambda_0$ of each source is then given as:

\begin{equation}
\lambda _0=(1+z)\lambda _e,
\end{equation}

\noindent where z is the redshift of the source and $\lambda_e$=6$\mu$m. Using the estimated value for $\lambda_0$, we apply an interpolation between the four WISE bands, to estimate the AB magnitude of each source, $m_{AB}$. We note that our sources are detected in all four WISE bands: 3.4, 4.6, 12 and 22 $\mu$m. For those sources that lie at very high redshift ($\gtrsim 3$) and their $\lambda _0$ falls at wavelengths longer than 22 $\mu$m an extrapolation is applied. The observed flux, $f_\nu(\nu_0)$, of the AGN is then given by the expression:

\begin{equation}
m_{AB}=-\frac{5}{2}log_{10}f_\nu(\nu_0)-48.6.
\end{equation}

\begin{figure}
\begin{center}
\includegraphics[height=1.1\columnwidth]{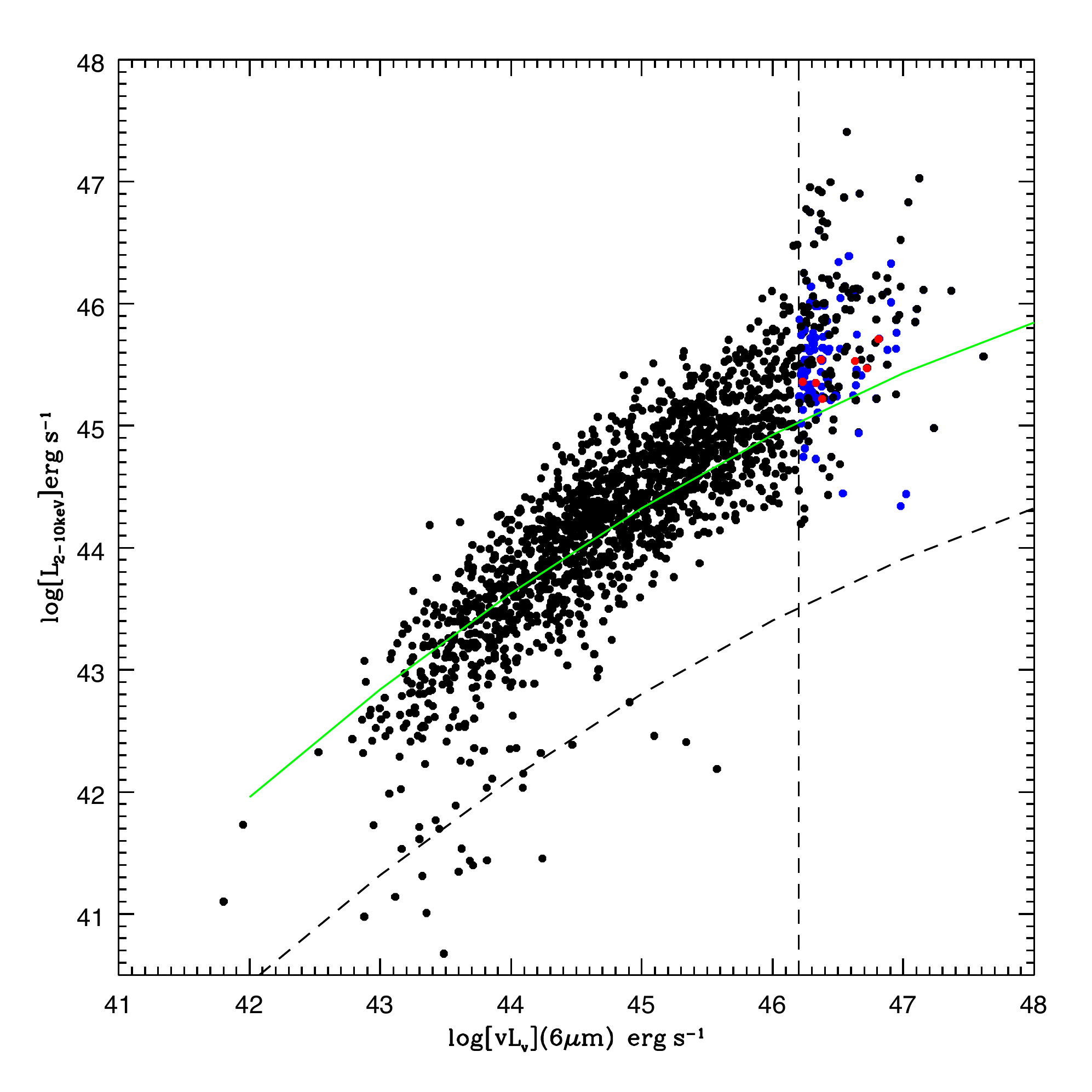}
\caption{2-10\,keV (observed) luminosity as a function of the 6 $\mu$m luminosity for the 2,834 common sources between the WISE catalogue and the 3XMM. The sources that belong to our luminous sample (see text) are shown in blue. The luminous and absorbed sources, presented in Table \ref {table_xspec_lum}, are shown in red. The solid vertical line indicates the luminosity cut for characterising a source as luminous. The green line is the X-ray to mid-IR relation \citep{Stern2015}. This relation has been chosen since it is appropriate fro a large range of luminosities, from Seyfert galaxies to luminous QSOs. The black-dashed line shows the luminosity-dependent average relation for Compton-thick (CT) AGN, assuming a 3\% fraction of reflected emission \citep{Fiore2009}.}
\label{fig_lx_vlv}
\end{center}
\end{figure}

\noindent Finally, the infrared luminosity is estimated as,

\begin{equation}
L_\nu=\frac{4\pi D^2_L}{1+z}f_\nu(\nu_0),
\end{equation}

\noindent where $D_L$ is the luminosity distance. The infrared luminosity distribution of our sample is presented in Fig. \ref {fig_lum_distrib}. The X-ray vs. IR luminosity for the 2,834 AGN, is presented in Fig. \ref{fig_lx_vlv}.

A luminosity cut at $\rm{log}\,\nu L_\nu\geq 46.2$\,erg\,s$^{-1}$, is applied to select only luminous sources. We choose this IR cut as it corresponds to $\rm{log}\,L_X \approx 45$\,erg\,s$^{-1}$ \citep{Stern2015} at the median redshift of our sample ($z=2.3$) and lies at the bright end of the X-ray luminosity function at $z>2$ \citep[e.g.,][]{Aird2010, Ueda2014, Georgakakis2015}. In addition, at these bright luminosities, the contamination from the star emission the AGN hosts to the W1 and W2 bands is negligible \citep[see Fig. 3 in][]{Mateos2015}. Since we shall also check the optical absorption of those sources found to be X-ray absorbed, we also exclude AGN that  do not have available optical spectrum from SDSS/BOSS \citep{Alam2015}. Our final sample consists of 65 AGN. Although we do not specifically look for type-1 AGN, our selection criteria bias our luminous sample towards type-1 sources. In particular, at high redshifts, SDSS selects, based on colours, Type-1 sources, i.e., QSOs.

\section{Analysis}
\subsection{X-ray absorption}

We fitted the X-ray spectra of our 65 sources by using Xspec v12.8
\citep{Arnaud1996}. We used Cash statistics \citep[implemented as C-stat in
Xspec][]{Cash1979} to the spectra binned to 1 count/bin, which has been proven to
recover the actual spectral parameters in the most accurate way even
for very low count spectra \citep[see for example][]{Krumpe2010b}.  All
the spectral models applied include photoeletric absorption fixed at
the Galactic value at the source coordinates given by
Leiden/Argentine/Bonn (LAB) Survey of Galactic HI \citep{Kalberla2005}.  We first fitted a power-law model absorbed by neutral
material totally or partially covering the central source (zwabs and
zpc in Xspec notation, respectively). A partial-covering absorbed
modelled this way can also represent a soft-scattered second power-law
component. We also checked for the presence of the Fe K$\alpha$ emission
line, the most common emission line observed in X-ray spectra of AGN,
by adding a gaussian component with centroid and width fixed at 6.4
keV and 100 eV rest-frame, respectively.

A source is considered as absorbed when the measured best-fit
intrinsic column density is higher than $10^{22}$\,cm$^{-2}$ and the
statistical lower limit (confidence limit 90\%) is N$_{H,low}>10^{21.5}$\,cm$^{-2}$. 
By applying the neutral absorption models, our spectral
fitting analysis reveals seven absorbed sources. The resulting
parameters for these seven sources are presented in Table \ref{table_xspec_lum}. Upper
limits for the Fe K$\alpha$ EW correspond to this component not being
required in the spectral fit. In the cases of sources d) and e), we
were forced to fix the photon index to 1.9 \citep[a typical value for
luminous AGN;][]{Corral2011}. The X-ray spectrum of source e) is
of too low quality to be able to constrain both the column density and
the photon index. In the case of source d), leaving it free results in
an unphysical value $<$ 1.3. Given this flat spectrum, we also tried a
power-law plus neutral reflection model, but the resulting spectral
parameters were unconstrained, probably due to the limited spectral
quality.

Significant X-ray absorption in type 1 AGN can be reconciled with AGN
unified models if this absorption comes from partially dust-free and
ionized material, which would have little effect on the optical
reddening. Ionized absorption from outflowing material has been
detected in the optical spectra of some AGN (Broad Absorption Line
QSOs; BALQSOs). The optical spectra for six out of our seven sources,
shown in Fig. 5 present evidence for broad [CIV] absorption
lines suggesting they could be classified as BALQSOs.

Page et al. (2011) presented a sample of five X-ray absorbed, broad
line QSOs, at $z\sim 2$, from ROSAT \citep{Page2001b}. Some of them show
evidence of [CIV] absorption lines in their ultraviolet spectra. Based
on their analysis, the X-ray spectra can be modelled successfully with
ionized absorbers, or with cold absorbers if they possess unusually
flat X-ray continuum shapes and unusual optical to X-ray spectral
energy distribution. They conclude that this population of sources,
represents the transition phase between obscured accretion and the
luminous QSO phase in the evolution of massive galaxies.

\begin{figure*}
     \begin{center}
           \subfigure[3XMMJ001232.0+052658]{
            \label{fig:first}
            \includegraphics[width=0.4\textwidth]{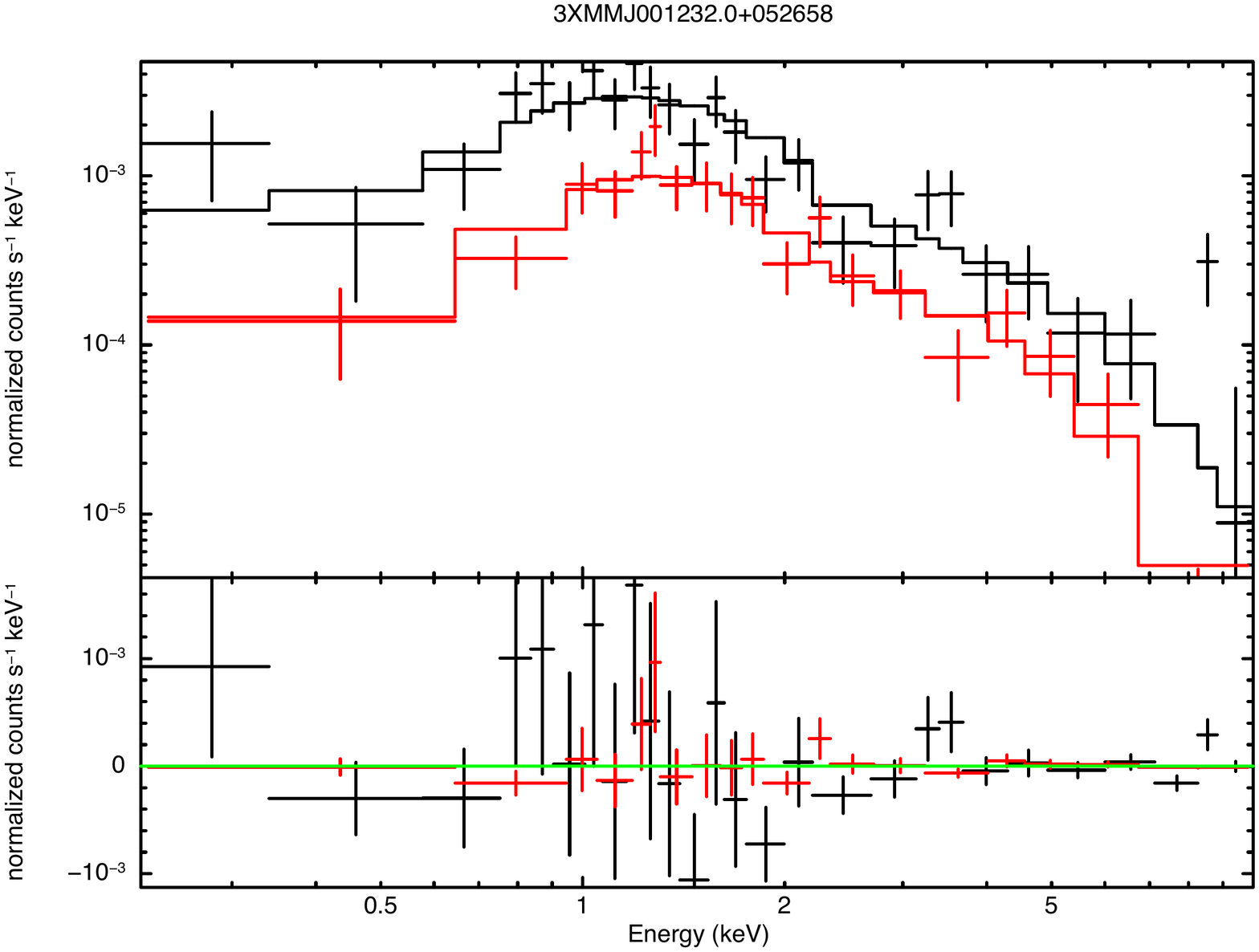}}
            \subfigure[3XMMJ153703.9+533219]{
            \label{fig:second}
            \includegraphics[width=0.4\textwidth]{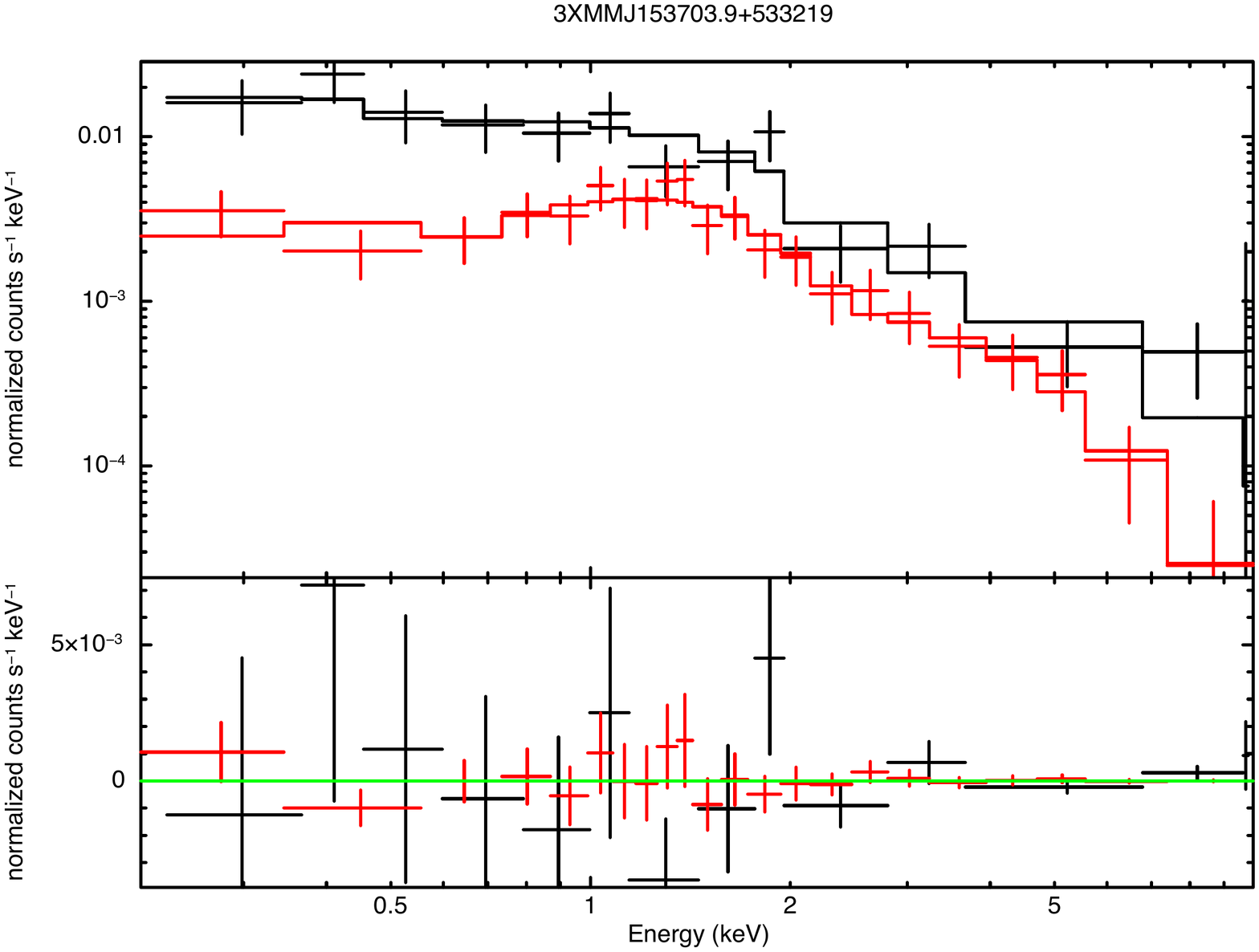}}\\
            \subfigure[3XMMJ090122.6+204446]{
            \label{fig:third}
            \includegraphics[width=0.4\textwidth]{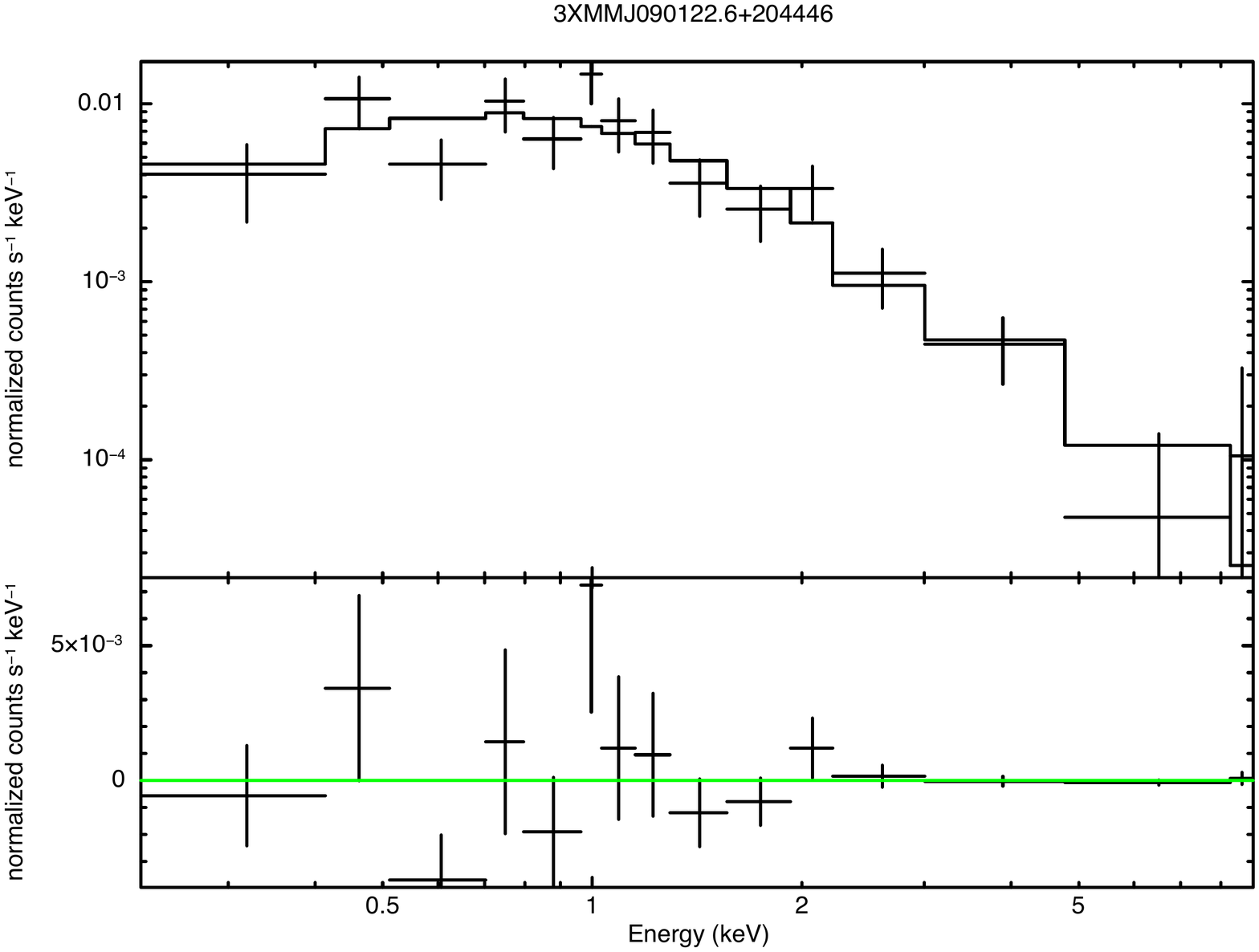}}
            \subfigure[3XMMJ120445.3+310609]{
            \label{fig:fourth}
            \includegraphics[width=0.4\textwidth]{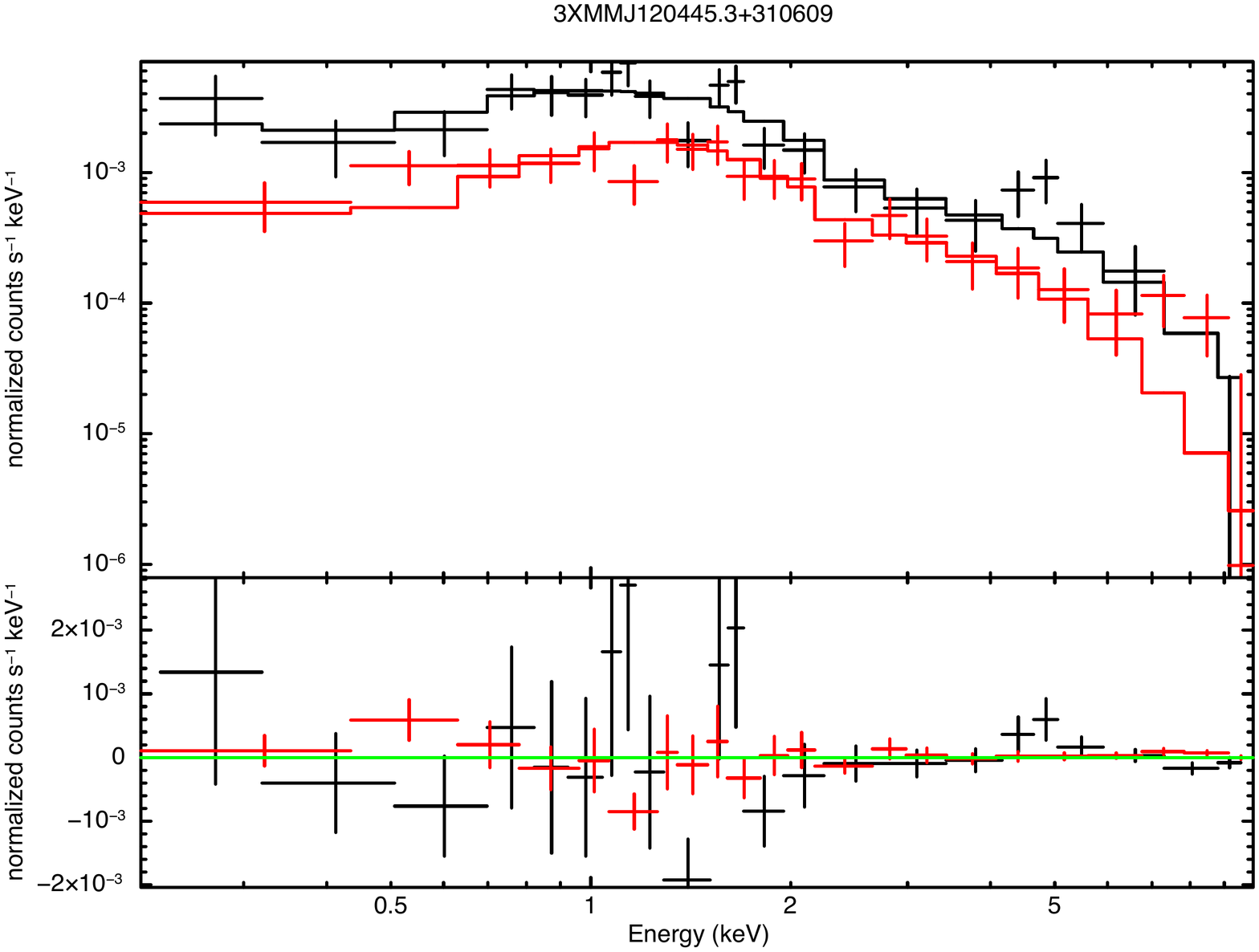}}
            \subfigure[3XMMJ114312.1+200346]{
            \label{fig:fifth}
            \includegraphics[width=0.4\textwidth]{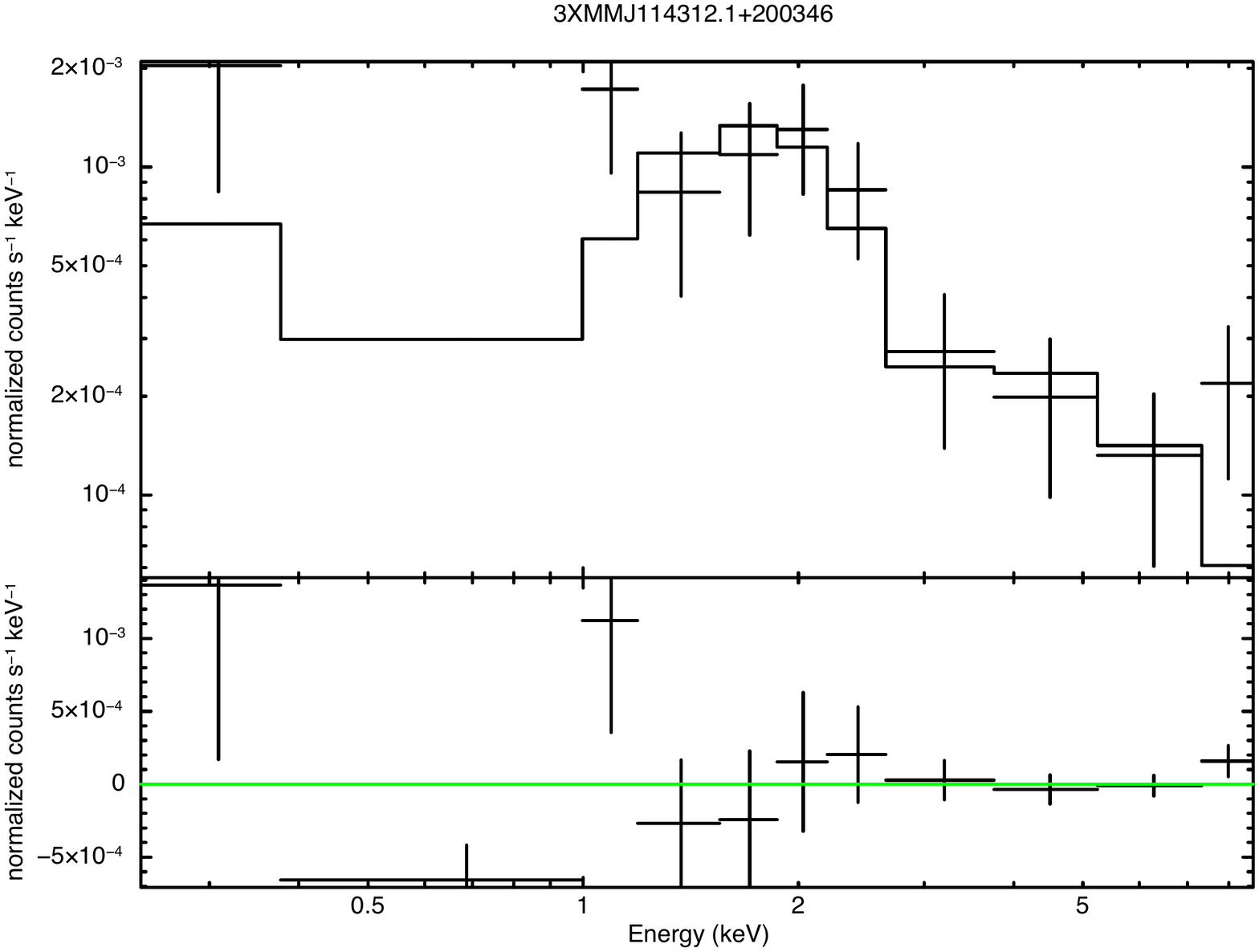}}
            \subfigure[3XMMJ024933.4-083454]{
            \label{fig:sixth}
            \includegraphics[width=0.4\textwidth]{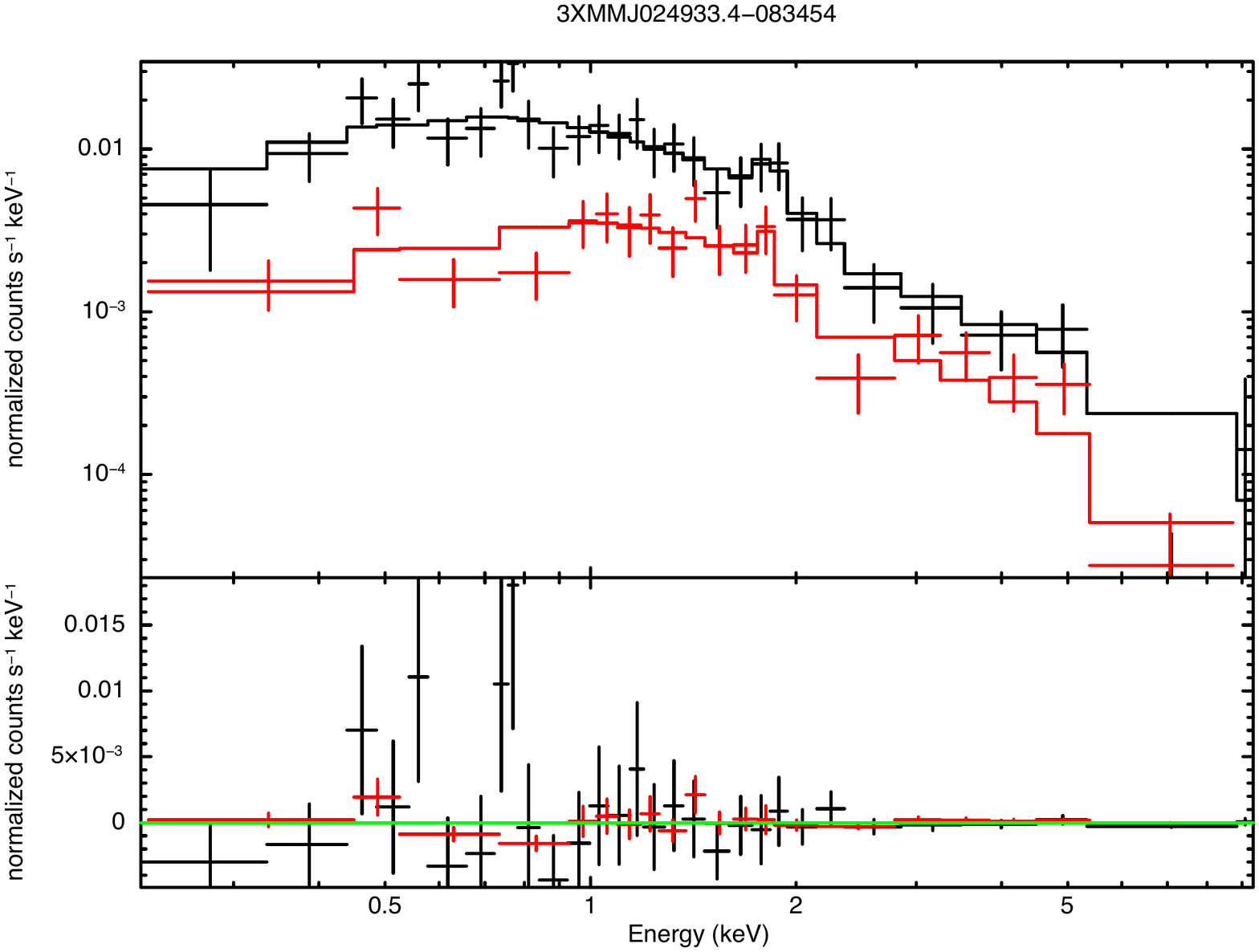}}
            \subfigure[3XMMJ112320.7+013747]{
            \label{fig:seventh}
            \includegraphics[width=0.4\textwidth]{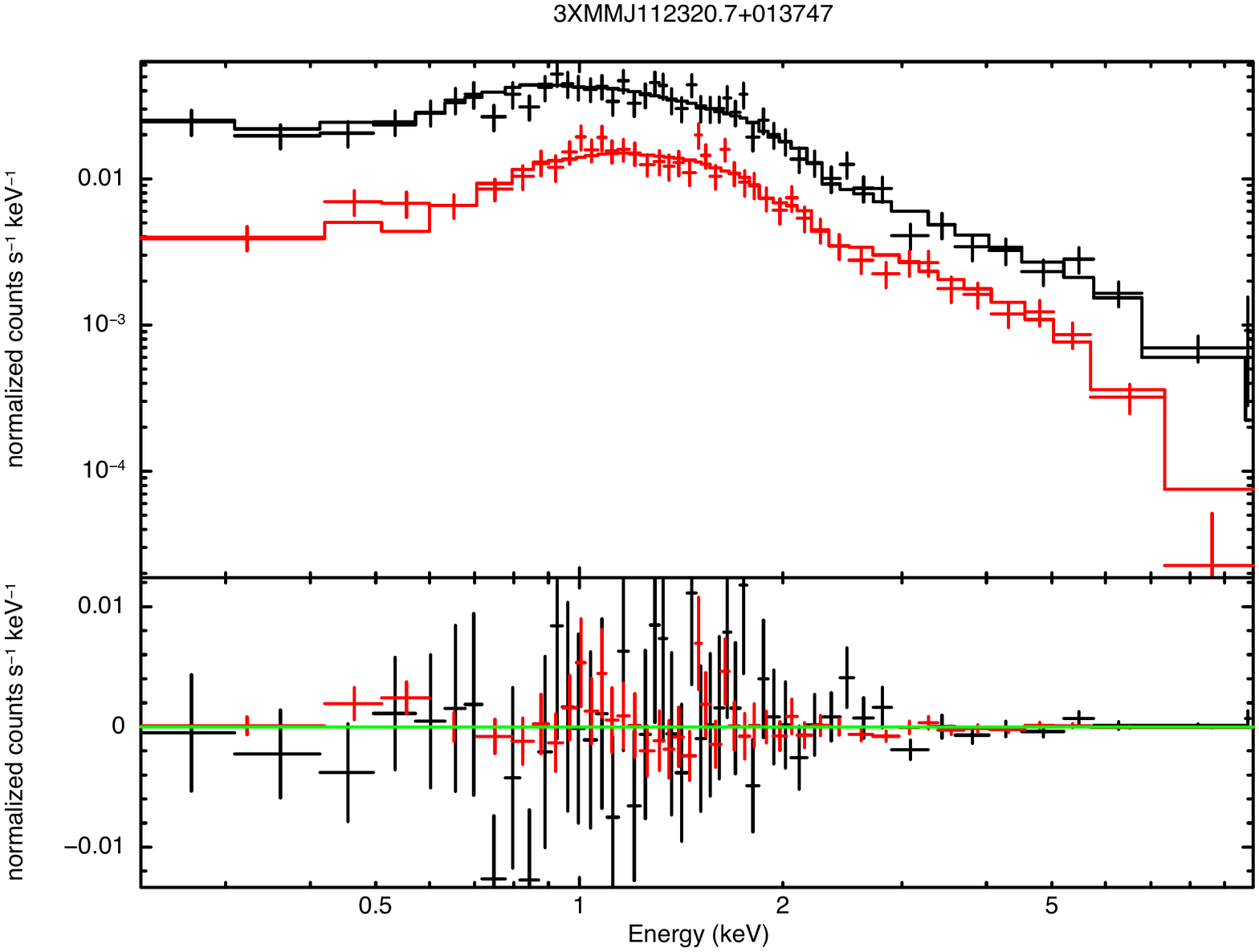}}
    \end{center}
    \caption{The X-ray spectra of our absorbed and luminous AGN. pn observations are presented in black, co-added MOS observations are presented in red, along with best-fit power-law models. The residuals are shown in the lower panels.}
   \label{fig_xspec}
\end{figure*}

To check whether the measured X-ray absorption for our seven sources could
come from ionized material, we also applied an ionized absorption
model (absori in Xpsec) to our luminous and absorbed sources. The
variable parameters of our ionized absorption model were the column
density and the ionization parameter $\xi$. $\xi$ represents the
ionization state of the asorbing material as: $\xi=L/nr^{2}$; where L is
the illuminating luminosity, n is the gas density, and r is the
distance from the illuminating source to the reprocessing gas. The
results are presented in Table \ref{table_xspec_lum}. The improvement with respect to the
neutral absorption model was estimated by using the likelihood ratio
test (l.r.t. in Table \ref{table_xspec_lum}), with values close to 1 meaning that the
neutral model is rejected in favor of the ionized one with high
significance. The l.r.t. test reveals that in five cases the ionized
model provides a better fit than the neutral model at the 1\,$\sigma$
confidence level (l.r.t.$>$ 0.7), so we consider the ionized absorption
model as the best-fit model in these cases. In all cases the NH values
derived from the ionized model are equal or larger than those derived
from the neutral model. The spectral fits corresponding to the
best-fit model for the seven luminous absorbed sources are shown in Fig. \ref{fig_xspec}. Finally, none of the remaining of our 65 luminous sources satisfies our obscuration criteria using the ionized absorption model.

\subsection{Optical absorption}

In this Section, we test whether the seven X-ray absorbed sources revealed by our XSPEC analysis show evidence of absorption based on their optical spectra or optical/near-IR colours.

\begin{figure*}
     \begin{center}
        \subfigure[3XMMJ001232.0+052658]{
            \label{fig:first}
            \includegraphics[width=0.45\textwidth]{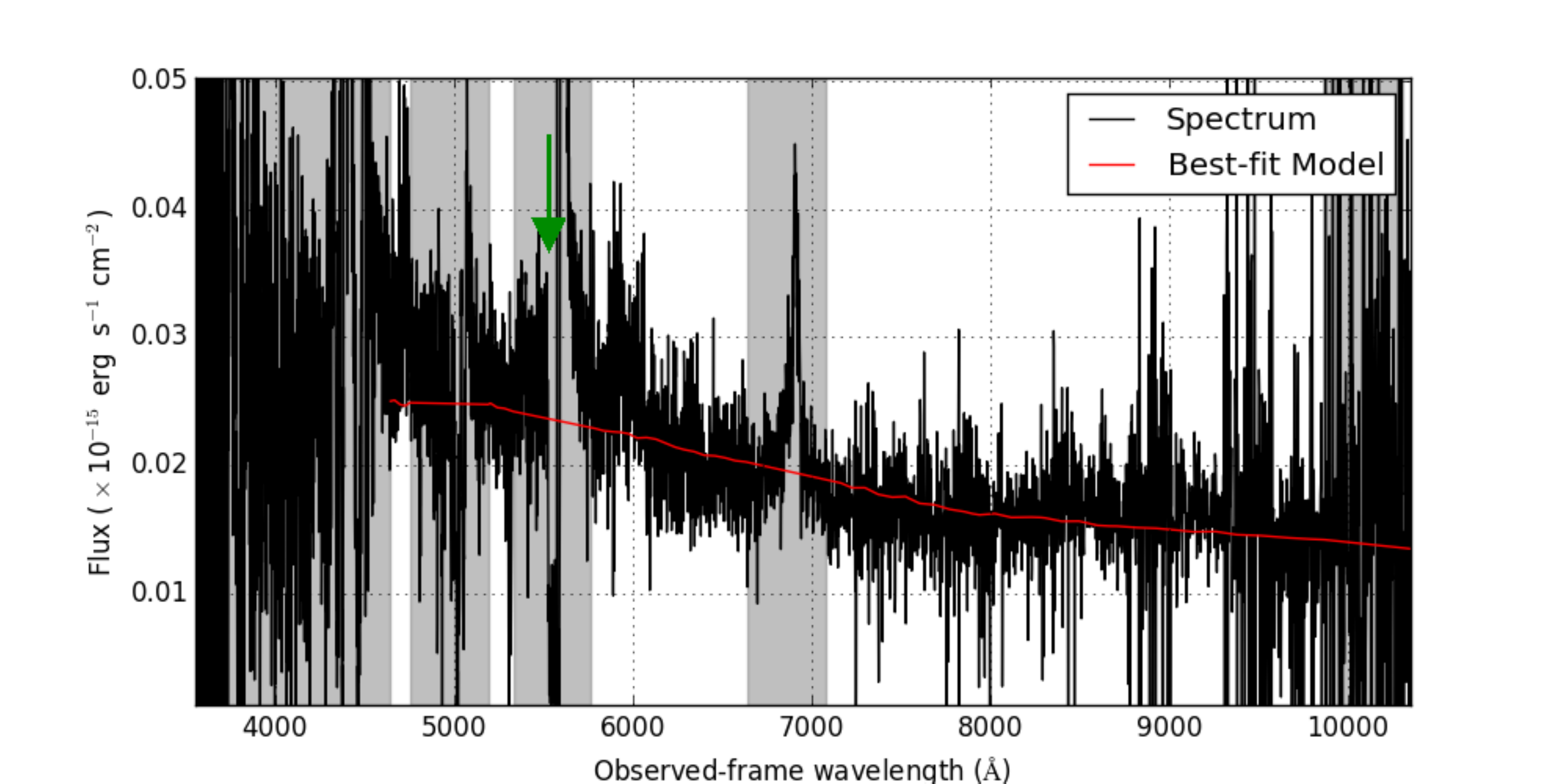}}            
          \subfigure[3XMMJ153703.9+533219]{
           \label{fig:second}
           \includegraphics[width=0.45\textwidth]{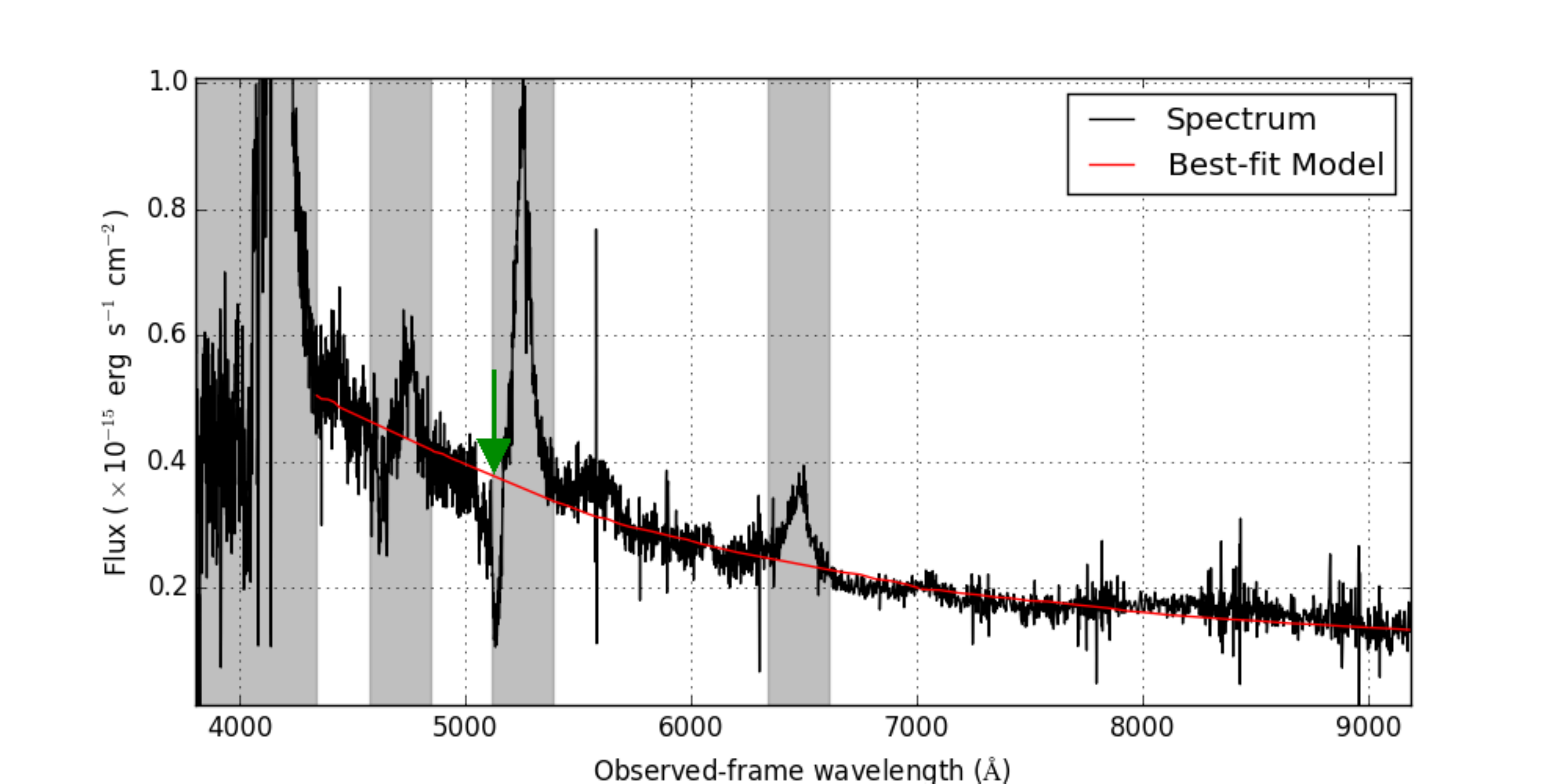}}\\                   
            \subfigure[3XMMJ090122.6+204446]{
             \label{fig:third}
              \includegraphics[width=0.45\textwidth]{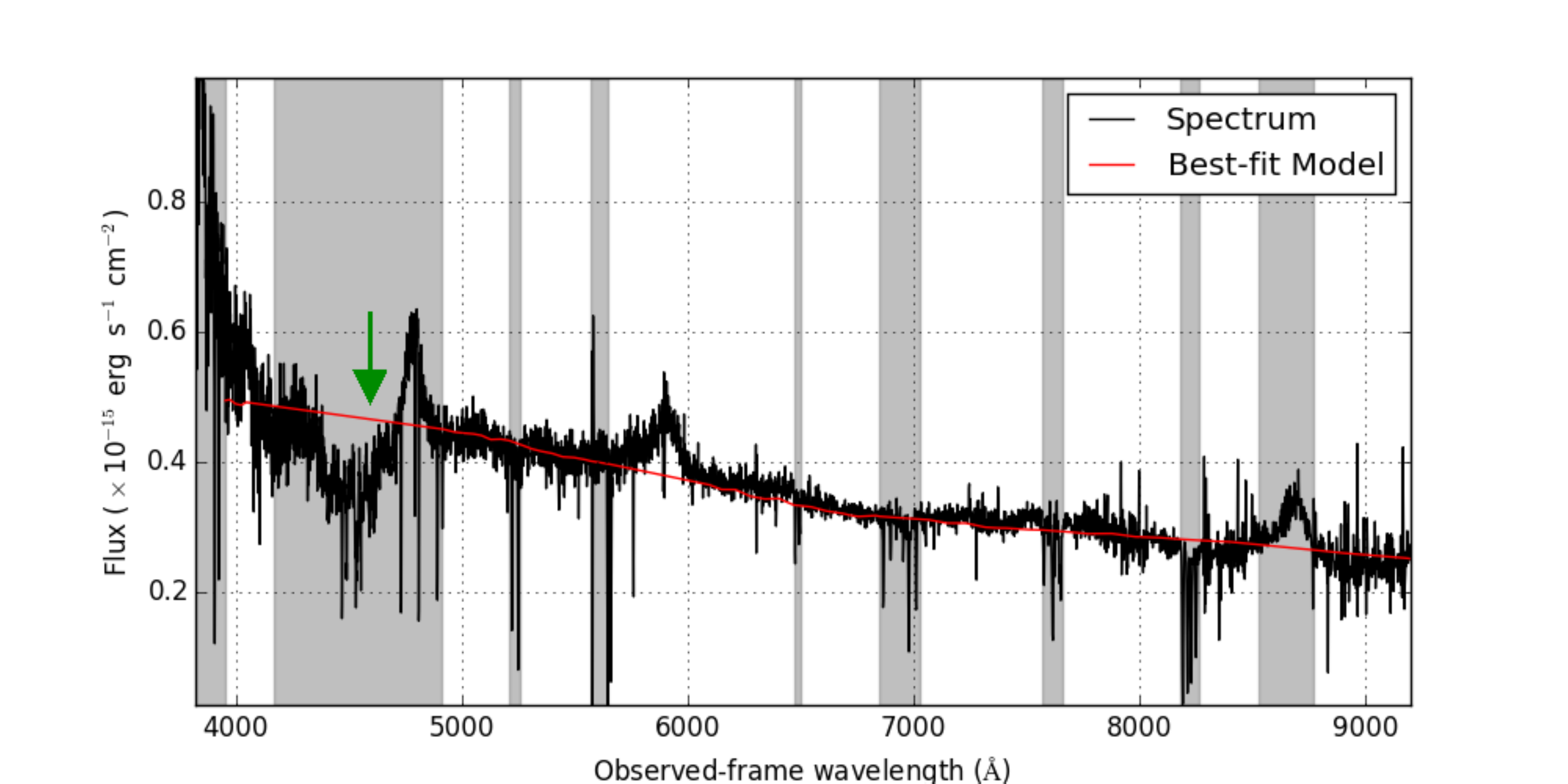}}                      
             \subfigure[3XMMJ120445.3+310609]{          
              \label{fig:fourth}
             \includegraphics[width=0.45\textwidth]{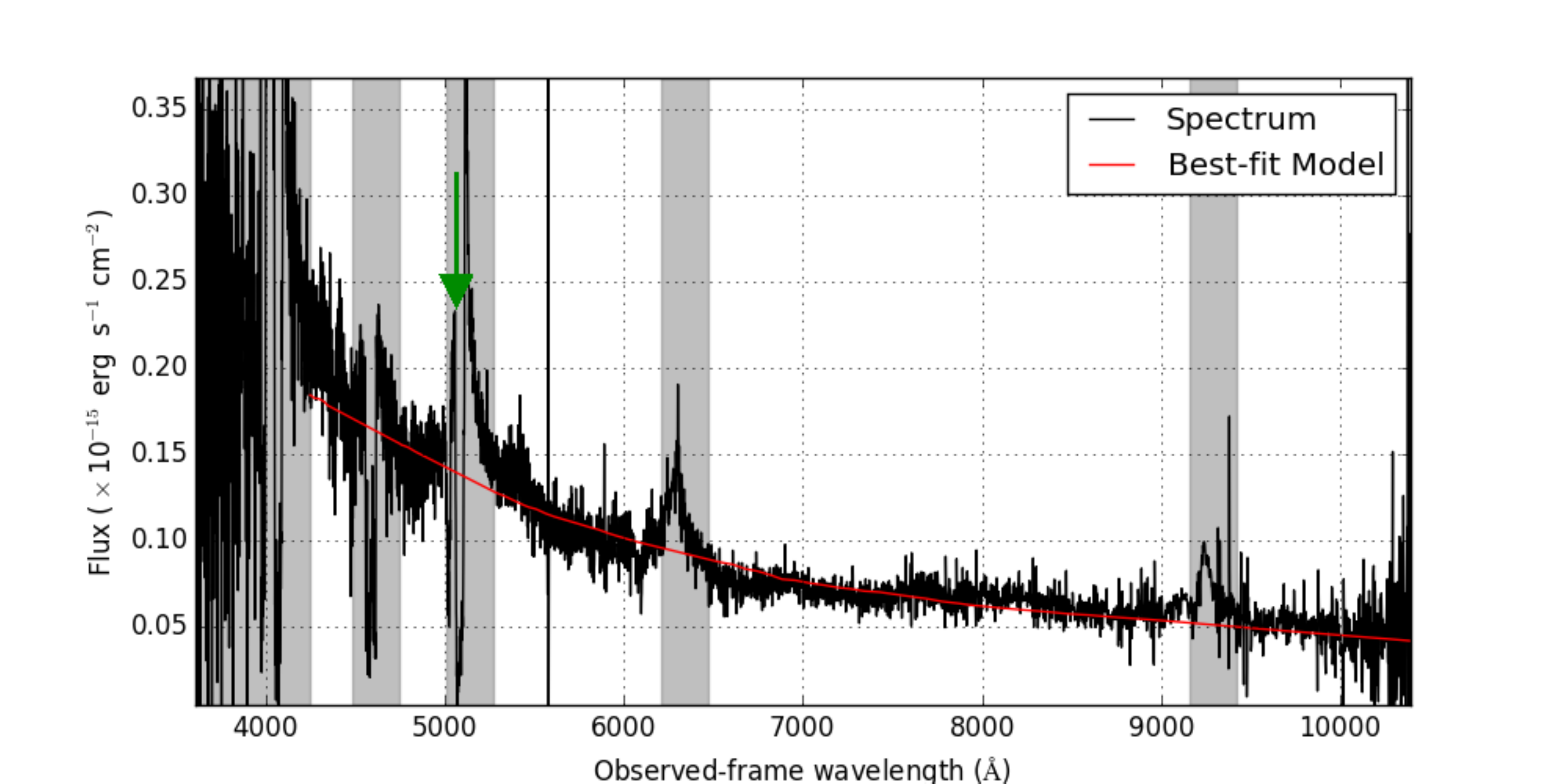}}\\           
             \subfigure[3XMMJ114312.1+200346]{          
              \label{fig:fitth}
             \includegraphics[width=0.45\textwidth]{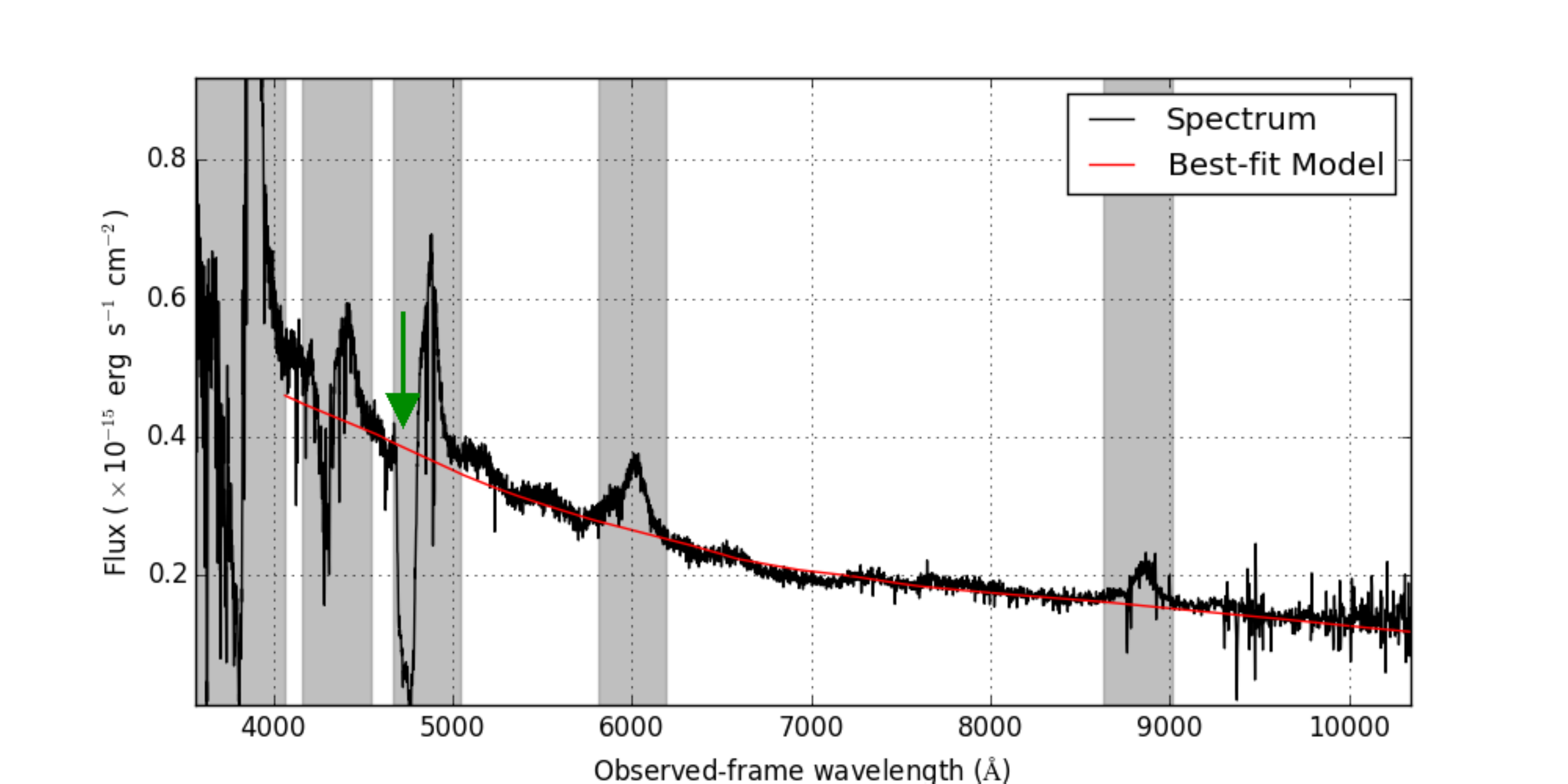}}      
             \subfigure[3XMMJ024933.4-083454]{          
              \label{fig:sixth}
             \includegraphics[width=0.45\textwidth]{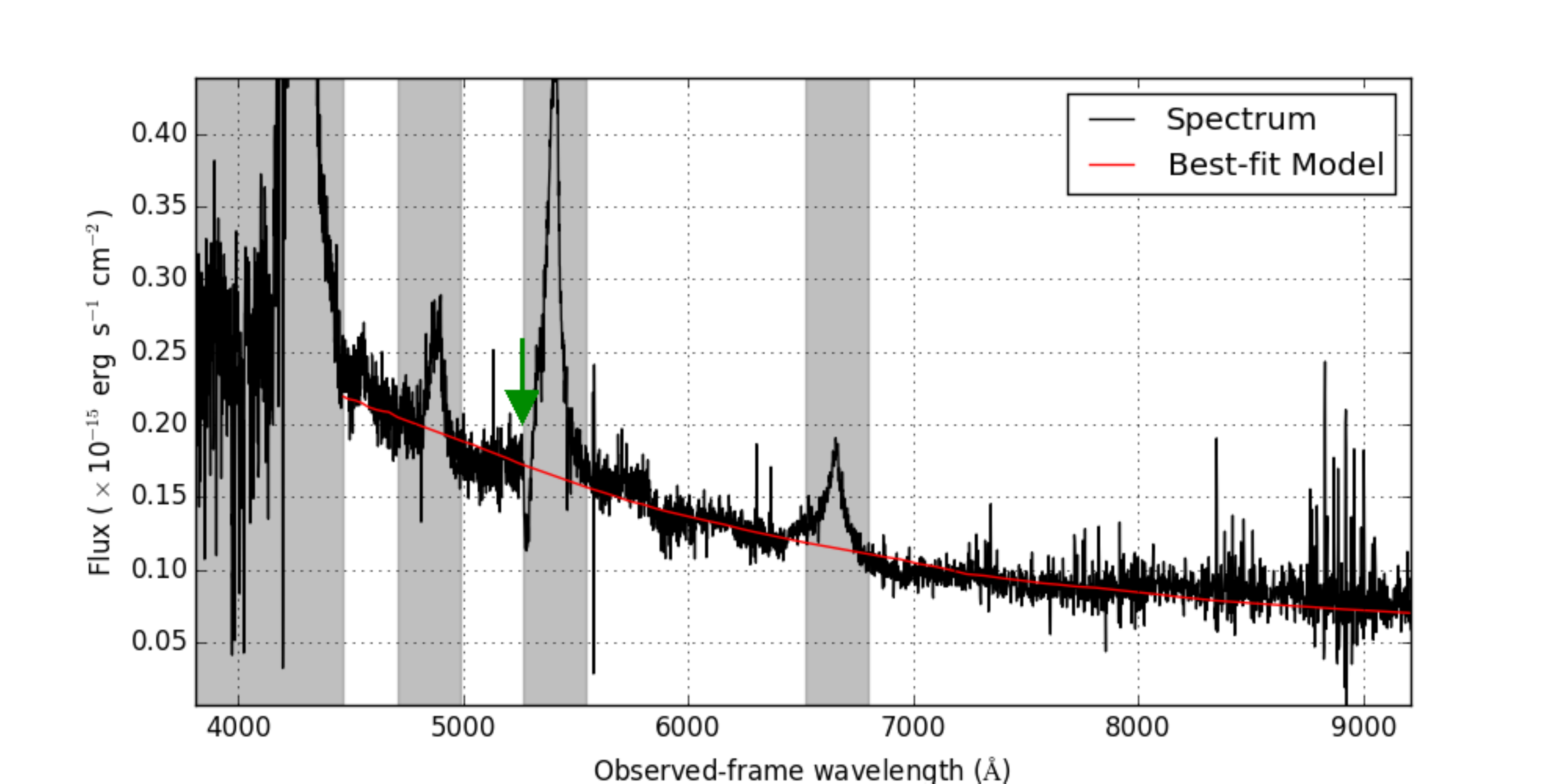}}\\   
             \subfigure[3XMMJ112320.7+013747]{          
              \label{fig:sixth}
             \includegraphics[width=0.45\textwidth]{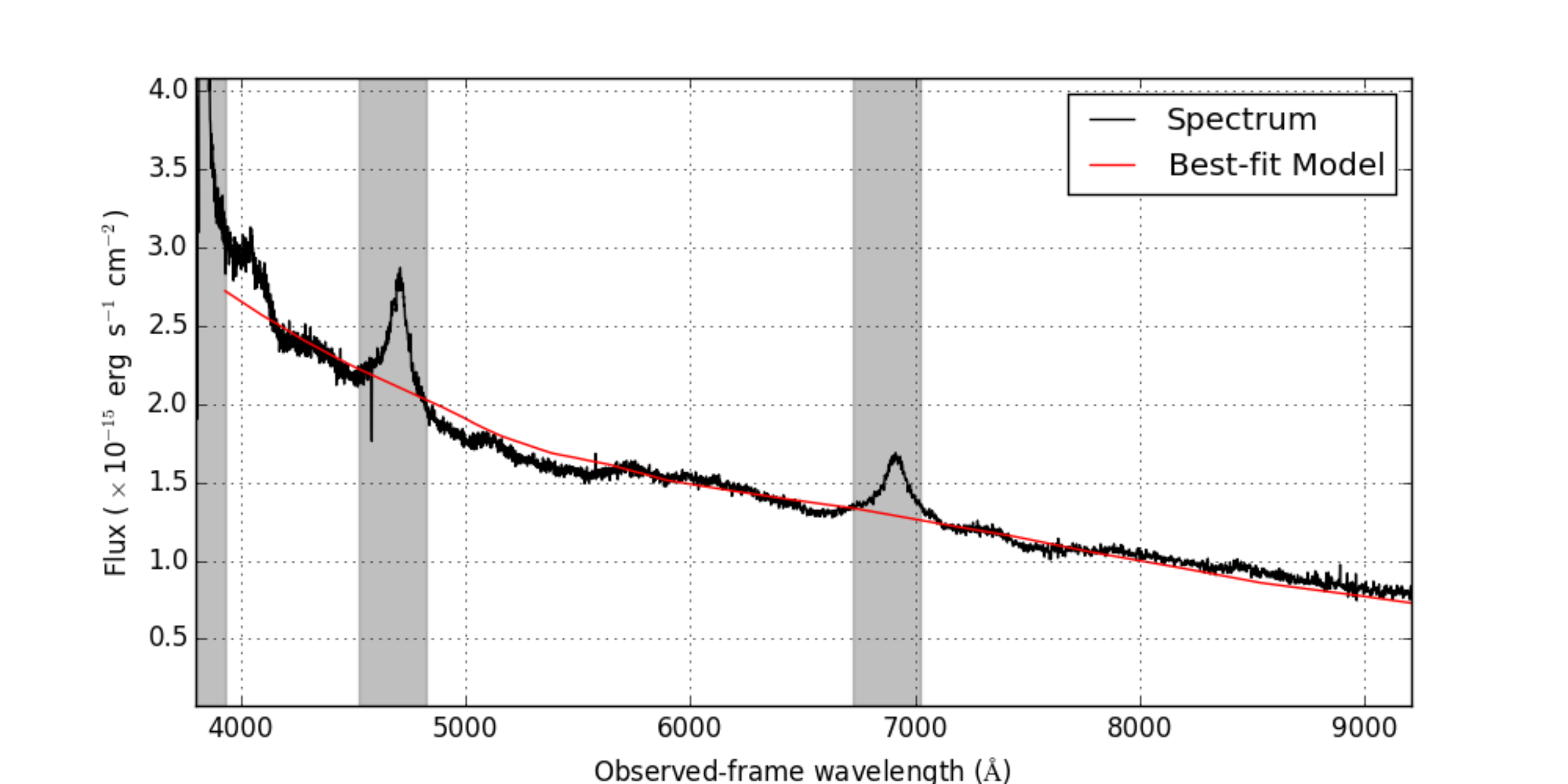}}\\         
                \end{center}
    \caption{Optical spectra of absorbed and luminous AGN. Regions ignored by the fits (see text) are shown as grey bands. A green arrow indicates the position of the [CIV] broad absorption line in each panel. Source 3XMMJ112320.7+013747 (panel-g) is not classified as BALQSO and there is no prominent absorption feature.}
   \label{fig_optical_spec}
\end{figure*}

For that purpose we use optical spectra obtained from the DR12 of the SDSS/BOSS and analyze the nuclear continuum of the selected AGN. All the spectra are corrected from Galactic extinction using the values from Leiden/Argentine/Bonn (LAB) Survey of Galactic HI \citep{Kalberla2005}. Since our objects are very bright no host galaxy subtraction is conducted. We fit the spectrum to a model of AGN emission, using the Richards QSO template \citep{Richards2006}, obscured using the Small Magellanic Cloud (SMC) extinction model from \citep{Gordon2003}. We use this model as it gives better results for modelling the extinction in AGN \citep{Hopkins2004}. The fits are performed using the CIAO's SHERPA fitting tool \citep{Freeman2001}. We allow the fits {\bf{to}} end with A$_{V}<$0 as some AGN are bluer than the Richards AGN template. The broad AGN emission lines are also masked out (as well as the absorption in the case of the Broad Absorption Line QSOs, BALQSOs; see next Section) as the extinction is measured using only the nuclear continuum. The optical spectra and the corresponding fits are presented in Fig. \ref{fig_optical_spec}. The A$_{V}$  values from the fits are shown in Table \ref{table_xray_opt_absorption}. The measured A$_{V}$ magnitude is far below A$_{V}$=2, used in previous studies to separate obscured and unobscured sources \citep{Caccianiga2008}. We caution however that our sources have an AGN dominated spectrum that peaks towards bluer colours. This A$_{V}$ value, using a Galactic dust-to-gas ratio, corresponds to N$_{H}$=4$\times$10$^{21}$\, cm$^{-2}$. Previous studies find that AGN do not follow the Galactic dust-to-gas ratio \citep{Maiolino2001}. In this case, assuming R$_V=3.1$ the corresponding N$_H$ value increases by about an order of magnitude.

Prompted by previous studies that find a correlation between the X-ray obscuration and the optical colour \citep[e.g.,][]{Civano2012} we apply the optical colour criteria of \cite{LaMassa2016} to our luminous sample of 65 AGN to select red sources. Using equations (1) and (2) in \cite{LaMassa2016} we estimate the values in the R bandpass. Our analysis reveals that five of our luminous sources are optically red (R-W1$>4$). One of them is also absorbed based on our X-ray spectra fitting, i.e., 3XMMJ001232.0+052658. \cite{Yan2013} showed that, at z$\leq 3$, type-2 AGN candidates can be identified using the following optical colour criteria: WISE $W1-W2>0.8, W2<15.2$ combined with $r-W2>6$ (Vega). Applying these criteria on our luminous sample we identify 5 sources. One of these sources is also X-ray absorbed, based on our XSPEC analysis. Table \ref{table_xray_opt_absorption} compares the X-ray absorption with the optical obscured criteria for our X-ray absorbed AGN.

\begin{table*}
\caption{The X-ray properties of our luminous (\rm{log}\,L$_{6\,\mu m}\ge 46.4$) and absorbed AGN. The intrinic X-ray luminosity (column 4) is estimated in the 2-10\,keV energy band. The $N_H$ values are shown in columns 6 and 11, for the neutral and ionized models, respectively. The likelihood ratio test (l.r.t) that compares the two models is shown in column 14. Column (7) presents the Covering Factor percentage (CF) and columns (9) and (14) present the C-stat values and the degrees of freedom (d.o.f.) for each model.}
\centering
\setlength{\tabcolsep}{0.3mm}
\begin{tabular}{cccc|ccccc|ccccc}
       \hline
 \hline
 
{3XMM ID} & {$\rm z$} & {$\nu$L$_{6\,\mu m}$}&{$\rm \log L_X$}  &{$\Gamma$}  & {$N_H$} & CF($\%$) & EW & C-stat/d.o.f. &{$\Gamma$}  & {$N_H$}&  $\xi$& cstat/dof & l.r.t.\\
 & &  ($\rm erg \,s^{-1}$) & ($\rm erg \,s^{-1}$) & &  ($10^{22}$\,cm$^{-2}$) &  & (eV) & & &($10^{22}$\,cm$^{-2}$) &&& \\
 \hline
 &&&& \multicolumn{5}{c}{Neutral absorption} & \multicolumn{5}{c}{Ionized absorption} \\
 (1) & (2) & (3) & (4) & (5) & (6) & (7) & (8) & (9) &(10)&(11)&(12)&(13)&(14) \\
       \hline
(a) 3XMMJ001232.0+052658 & 2.63 & 46.81 & 45.65    & $1.9^{+0.3}_{-0.3}$ & $9^{+3}_{-4}$   & $94^{+6}_{-8}$    & $<$ 200      &  372/398 & $1.8^{+0.2}_{-0.2}$    &  $29^{+11}_{-23}$ & $<$1600           & 374/398   & 0.24\\
(b) 3XMMJ153703.9+533219 & 2.40 & 46.37 & 45.46    & $1.9^{+0.4}_{-0.4}$ & $12^{+12}_{-8}$  & $60^{+20}_{-40}$  & $<$ 300 &  245/294& $1.8^{+0.1}_{-0.2}$ &  $18^{+18}_{-16}$ & $1700^{+2500}_{-900}$  & 246/294   & 0.77\\
(c) 3XMMJ090122.6+204446 & 2.09 & 46.63 & 45.55    & $1.8^{+0.5}_{-0.4}$ & $0.9^{+1.0}_{-0.7}$   &    & $<$ 600  &  104/135 & $1.9^{+0.5}_{-0.4}$    &  $1.1^{+12}_{-0.9}$ & $<$ 950          & 104/134   & 0.13   \\
(d) 3XMMJ120445.3+310609 & 2.32 & 46.23 & 45.06    & $1.9^{f}$         & $2.2^{+1.5}_{-0.9}$   &       & $110^{+320}_{-110}$ &  375/395&  $1.9^{f}$ &$13^{+11}_{-5}$    & $440^{+610}_{-250}$&365/394  & 1.0\\
(e) 3XMMJ114312.1+200346 & 2.18 & 46.38 & 44.92    & $1.9^{f}$         & $21^{+15}_{-8}$      &         & $<$700         &  100/107 &  $1.9^{f}$ & $100^{+80}_{-50}$  & $2100^{+2200}_{-1400}$ &  97/106   & 0.97\\
(f) 3XMMJ024933.4-083454 & 2.49 & 46.33 & 45.37    & $1.9^{+0.2}_{-0.2}$ & $1.0^{+0.6}_{-0.6}$   &        & $460^{+380}_{-340}$ &  348/388 &$1.9^{+0.2}_{-0.2}$ & $1.3^{+3.5}_{-0.7}$ & $<$300           &  345/387  & 0.72\\
(g) 3XMMJ112320.7+013747 & 1.48 & 46.72 & 45.40    & $2.0^{+0.1}_{-0.1}$ & $2.7^{+0.7}_{-0.6}$ & $83^{+4}_{-5}$  & $<$ 140       &  901/1050& $1.9^{=0.1}_{-0.1}$ &$3.5^{1.2}_{-1.1}$  & $100^{+60}_{-50}$    & 907/1050  & 1.0 \\

\hline
\label{table_xspec_lum}
\end{tabular}
\end{table*}

\begin{table}
\caption{Comparison of X-ray obscuration with optical obscuration. For the third optical criterion, i.e., $r-W2>6$, other criteria have to be satisfied, too (see text for more details). Sources that satisfy any optical obscuration criterion are shown in bold.}
\centering
\setlength{\tabcolsep}{1.5mm}
\begin{tabular}{cccc}
       \hline
 \hline
{3XMM ID} & A$_{V}$  & $R-W1$ & $r-W2$ \\
       \hline
    \\
\bf{3XMMJ001232.0+052658} &$0.17^{+0.05}_{-0.10}$ & \bf{4.78}& \bf{6.16}\\
3XMMJ153703.9+533219  &$-0.19^{+0.05}_{-0.01}$ &3.15 & 4.34  \\
3XMMJ090122.6+204446 &$0.19^{+0.01}_{-0.04}$ &3.45 & 4.76  \\
3XMMJ120445.3+310609 &$-0.09^{+0.06}_{-0.02}$ &3.28 & 4.57  \\
3XMMJ114312.1+200346  &$0.06^{+0.04}_{-0.03}$ &3.38 & 4.96  \\
3XMMJ024933.4-083454  &$-0.03^{+0.02}_{-0.01}$ & 2.96& 4.03  \\
3XMMJ112320.7+013747  &$0.02^{+0.03}_{-0.01}$  &3.25 &  5.12\\
\hline
\label{table_xray_opt_absorption}
\end{tabular}
\end{table}

\begin{table}
\caption{The balnicity index (BI) and the less stringent absorption index (AI), for five out of seven absorbed sources that are classified as BALQSOs in the SDSS (DR12) catalogue. }
\centering
\setlength{\tabcolsep}{1.5mm}
\begin{tabular}{ccc}
       \hline
 \hline
{3XMM ID} & BI  & AI   \\
& (Km\,s$^{-1}$) &  (Km\,s$^{-1}$) \\
       \hline
    \\
3XMMJ001232.0+052658 & 0.0 & $3618.7\pm 194.3$\\
3XMMJ153703.9+533219  &$1879.0\pm 8.4$ &$2295.9\pm 15.2$   \\
3XMMJ090122.6+204446 &$1624.1\pm23.4$ &$2111.4\pm37.5$  \\
3XMMJ120445.3+310609  & 0.0 & $3001.5\pm 18.9 $  \\
3XMMJ114312.1+200346  &$4064.1\pm2.3$ &$5173.2\pm3.6$   \\

\hline
\label{table_balnicity}
\end{tabular}
\end{table}

\section{Summary and Discussion}

We studied the X-ray properties of luminous IR-selected AGN by cross-matching the 3XMM catalogue with WISE selected AGN. Our analysis revealed 65 AGN with spectroscopic redshift that are luminous ($\rm{log}\,\nu L_\nu\geq 46.2$\,erg\,s$^{-1}$) and seven of them are also absorbed (log\,N$_H>$\,22 cm$^{-2}$, N$_{H,low}>10^{21.5}$\,cm$^{-2}$), based on our X-ray spectra fitting using XSPEC.  Our analysis suggests that $\approx 10\%$, i.e., seven out of 65 sources, of Type 1 AGN are obscured in X-rays.

We then tested whether our X-ray absorbed sources show obscuration in their optical spectra. Fitting the available SDSS/BOSS optical spectra we find no optical obscuration, based on their optical continuum. However, five out of the seven sources are classified as Broad Absorption Line (BAL) quasars in the DR12 SDSS catalogue \footnote{\textrm{https://data.sdss.org/datamodel/files/BOSS$\_$QSO/DR12Q/DR12Q$\_$BAL.html}}. Source (f) is the only one not classified as BALQSO while source (g) does not have the necessary wavelength coverage as the [CIV] line falls just bellow the 4000 Angstroms. Existence of absorption lines while there is no absorption in the optical continuum has been previously observed in the case of BAL quasars. These systems show no or little dust extinction in the optical/UV, but with broad, blueshifted absorption lines \citep{Barger2002}. They are thought to be intrinsically normal quasars, covered by a high column density of dust-free gas that is responsible for absorption in the X-rays and absorption lines in the optical/UV \citep{Brandt2000}. Table \ref{table_balnicity} presents the balnicity index \citep[BI;][]{Weymann1991} as well as the alternative, less stringent criterion to classify an object as BALQSO, the absorption index \citep[AI;][]{Hall2002, Trump2006} for our five BALQSOs. Both indices measure the strength of the C\,IV and Mg\,II absorption lines to separate the non-BALQSOs from the BALQSOs. A positive BI or AI indicates the source is a BALQSO. The BI index ($\rm{BI}=-\int_{25,000}^{3,000}[1-f(V)/0.9]CdV$) measures the equivalent width of the continuous troughs of absorption but, for example, ignores absorption $<$3000 km\,s$^{-1}$. The AI index ($\rm{AI}=\int_{0}^{25,000}[1-f(V)/0.9]CdV$) includes the contribution of absorption troughs with outflow velocities below 3000 km\,s$^{-1}$ and requires a minimum absorption trough width of only 1000 km\,s$^{-1}$ to yield a positive AI \citep{Page2016}. For both indices,  f(V) is the normalized flux as a function of velocity displacement from the line center and C is set to zero. It is set to 1.0 whenever the quantity in brackets has been continuously positive over an interval of 2000 km\,s$^{-1}$   \citep[for more information see Appendix in][]{Weymann1991, Hall2002}. In the remaining 58 sources of our luminous sample of 65 AGN, we find one more source that is classified as BAL, by SDSS. The column density of this source (3XMMJ154359.4+535902) falls marginally below our criterion to be characterized as absorbed (log\,N$_H=$\,21.9$\pm0.1${cm$^{-2}$). An ionized model, though, provides a better fit to its spectrum. In this case the column density is log\,N$_H=$\,22.7$\pm^{+1.0}_{-0.3}${cm$^{-2}$.



Recently, \cite{Page2016} studied a sample of six X-ray selected BALQSOs from the XMM-{\it{Newton}} Wide Angle Survey and found evidence for absorption in the X-ray spectra of all six objects. \cite{Streblyanska2010} studied a sample of 39 BALQSOs. 36\% of their sample presents low levels of obscuration when fitted using a neutral absorption model. However, when they apply a ionized absorption model, 90\% of their objects are absorbed. Ionized X-ray absorption may not be a better fit to the X-ray spectra, but it allows a simpler explanation to the optical/UV/X-ray spectrum without invoking special characteristics.


Previous studies have found that red AGN tend to be associated with higher levels of X-ray obscuration, based on their HRs \citep[e.g.,][]{Brusa2005, Fiore2009, Civano2012}. Applying optical to mid-IR colour criteria, i.e., R$-$W1, to select red sources in our luminous sample, revealed that five out of 65 sources are optically red, but only one of them was also absorbed in X-rays. \cite{LaMassa2016} used X-ray sources in the Stripe 82X, observed with the  XMM-Newton and Chandra missions. Their sample consists mostly of broad-line AGN at $z>1$. They find that their reddened AGN (R$-$W1$>$4) do not have higher HRs than the bluer AGN. Their percentage of X-ray obscured AGN among their red sources is in agreement with our estimations. However, our sample as well as that of Stripe 82X \citep{LaMassa2016} consist of broad-line AGN in contrast to the narrow-line sources used in  \cite{Brusa2005}, \cite{Fiore2009} and \cite{Civano2012}.

Moderate absorption in Type-1 AGN has also been reported in previous works. \cite{Wilkes2005} used Chandra and XMM-Newton observations to study a sample of five red, 2MASS AGN ($J-K_s>2$). They find that substantial absorbing material ($\sim 10^{22}$\,cm$^{-2}$) is present in three of these sources . According to their analysis, these sources are X-ray hard most probably because of absorption. They suggest that the intermediate level of obscuration they find is indicative of outflowing winds above a disk/torus. This conclusion is in agreement with that of \cite{Brusa2015}. In that work they presented X-shooter@VLT observations of a sample of 10, luminous, albeit Type-2, X-ray obscured QSOs at z$\sim$1.5 from the XMM-COSMOS survey. They conclude that the existence of strong outflows, in these sources, has an AGN origin. These objects are expected to be caught in the transitioning phase from starburst to AGN dominated systems. Banerji et al. (2013) used $\it{Herschel}$ and $\it{XMM-Newton}$ observations to study the broad-line Type 1 QSO, named ULASJ1234+0907. This is the reddest broad-line Type 1 quasar known, with (i-K)$_{AB}>7.1$. The QSO lies at high redshift ($z=2.503$) and has a hard X-ray luminosity of L$_{2-10\,keV}=1.3\times 10^{45}$\,erg\,s$^{-1}$. They measure a column density of N$_H=\,9.0\times 10^{21}$\,cm$^{-2}$, that is in accordance with the obscuration level found in our luminous and obscured sources. They conclude that their QSO is at the peak epoch of galaxy formation, transitioning from a starburst to optical quasar via a dusty quasar phase. \cite{Bischetti2017} used a sample of WISE/SDSS selected hyper-luminous ($L_{Bol}>10^{47}$\,erg\,s$^{-1}$) broad-line quasars at $z\approx1.5-5$. Five of their sources show prominent [OIII] emission lines, revealing the presence of powerful ionised outflows. Their results suggest that at high luminosities AGN are very efficient in pushing outwards large amounts of ionized gas.


In this paper, in accordance with previous works we find that $\approx 10\%$ of luminous, high redshift AGN with broad emission lines present moderate levels of X-ray absorption, i.e., N$_H$$\sim 10^{22}$\, cm$^{-2}$. Recent works using different selection criteria have found several samples of luminous AGN with evidence for absorbing and outflowing material, to which we have added our own sample of X-ray absorbed MIR-luminous sources. Those are the characteristics expected for the putative population of QSOs emerging from the obscured phase and blowing away the surrounding material in a galaxy-AGN co-evolution scenario.


\section{acknowledgmenets}

We thank the anonymous referee for helpful comments and Prof. M. Brusa for useful discussions. The research leading to these results has received funding from the European Union's Horizon 2020 Programme under the AHEAD project (grant agreement n. 654215). GM acknowledges financial support from the AHEAD project that is funded by the European Union as Research and Innovation Action under Grant No: 654215. FJC and IO-P acknowledge financial support through grant AYA2015-64346-C2-1-P (MINECO/FEDER). SM acknowledges financial support by the Spanish
Ministry of Economy and Competitiveness through grant AYA2016-76730-P. This research has made use of data obtained from the 3XMM XMM-Newton serendipitous source catalogue compiled by the 10 institutes of the XMM-Newton Survey Science Centre selected by ESA. Funding for SDSS-III has been provided by the Alfred P. Sloan Foundation, the Participating Institutions, the National Science Foundation, and the U.S. Department of Energy Office of Science. The SDSS-III web site is http://www.sdss3.org/. SDSS-III is managed by the Astrophysical Research Consortium for the Participating Institutions of the SDSS-III Collaboration including the University of Arizona, the Brazilian Participation Group, Brookhaven National Laboratory, Carnegie Mellon University, University of Florida, the French Participation Group, the German Participation Group, Harvard University, the Instituto de Astrofisica de Canarias, the Michigan State/Notre Dame/JINA Participation Group, Johns Hopkins University, Lawrence Berkeley National Laboratory, Max Planck Institute for Astrophysics, Max Planck Institute for Extraterrestrial Physics, New Mexico State University, New York University, Ohio State University, Pennsylvania State University, University of Portsmouth, Princeton University, the Spanish Participation Group, University of Tokyo, University of Utah, Vanderbilt University, University of Virginia, University of Washington, and Yale University.

\bibliography{mybib}{}

\bibliographystyle{mn2e}

\end{document}